\documentclass[floats,floatfix,showpacs,amssymb,prd,twocolumn,superscriptaddress,nofootinbib,nolongbibliography,reprint]{revtex4-2}

\usepackage{graphicx,epsf, epsfig, amssymb}
\usepackage{bm}
\usepackage{color}
\usepackage[breaklinks,bookmarksopen=true]{hyperref}
\usepackage{amsfonts,amsmath,amssymb,mathrsfs,gensymb}
\usepackage{natbib}
\usepackage{prd_macros}
\usepackage{array}
\usepackage{multirow}
\usepackage{rotating,array}
\usepackage[normalem]{ulem}
\usepackage{ulem}
\usepackage{dcolumn}
\usepackage{braket}
\usepackage{appendix}
\usepackage{subcaption}
\usepackage{comment}
\usepackage[dvipsnames, usenames]{xcolor}

\definecolor{linkcolor}{rgb}{0.0,0.3,0.5}
\usepackage[all]{hypcap}
\usepackage[T1]{fontenc}
\usepackage[utf8]{inputenc}
\usepackage{tabularx}

\usepackage [english]{babel}
\usepackage [autostyle, english = american]{csquotes}
\MakeOuterQuote{"}

\newcommand{\dallas}{\affiliation{Department of Physics, The University of Texas at Dallas, Richardson, Texas 75080, USA}}

\newcommand\orcid[1]{\href{https://orcid.org/#1}{$\!$\includegraphics[scale=0.006]{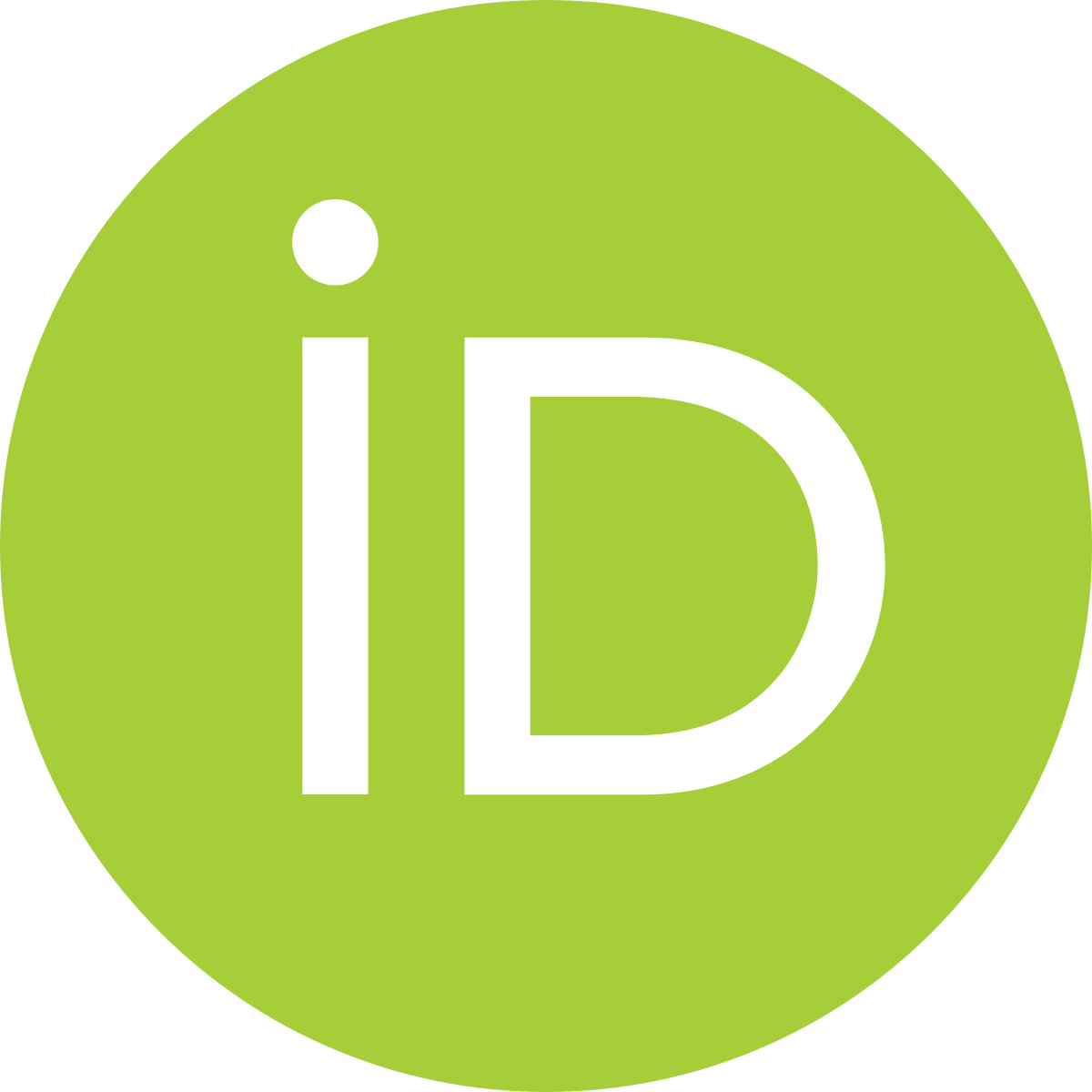} $\!\!$}}

\begin{document}


\title{Two regimes of tidal-stream circularization by supermassive black holes}

\author{Joseph Rossi \orcid{0000-0001-6712-1788}}
\email{jxr121830@gmail.com} 
\dallas

\author{Juan Servin}
\email{juanservin123@gmail.com}
\dallas

\author{Michael Kesden \orcid{0000-0002-5987-1471}}
\email{kesden@utdallas.edu}
\dallas

\date{\today}

\begin{abstract}
Stars that approach a supermassive black hole (SMBH) too closely can be disrupted by the tidal gravitational field of the SMBH.  The resulting debris forms a tidal stream orbiting the SMBH which can collide with itself due to relativistic apsidal precession. These self-collisions dissipate energy, causing the stream to circularize.  We perform kinematic simulations of these stream self-collisions to estimate the efficiency of this circularization as a function of SMBH mass $M_\bullet$ and penetration factor $\beta$, the ratio of the tidal radius to the pericenter distance.  We uncover two distinct regimes depending on whether the time $t_c$ at which the most tightly bound debris circularizes is greater or less than the time $t_{\rm fb}$ at which the mass fallback rate peaks.  The bolometric light curve of energy dissipated in the stream self-collisions has a single peak at $t > t_{\rm fb}$ in the slow circularization regime ($t_c > t_{\rm fb}$), but two peaks (one at $t < t_{\rm fb}$ and a second at $t_{\rm fb}$) in the fast circularization regime ($t_c < t_{\rm fb}$).  Tidal streams will circularize in the slow (fast) regime for apsidal precession angles less (greater) than 0.2 radians which occur for $\beta \lesssim (\gtrsim) (M_\bullet/10^6M_\odot)^{-2/3}$.  The observation of prominent double peaks in bolometric TDE light curves near the transition between these two regimes would strongly support our model of tidal-stream kinematics.
\end{abstract}

\maketitle

\section{Introduction} \label{S:Intro}

Tidal disruption events (TDEs) occur when stars approach closely enough to a supermassive black hole (SMBH) that the tidal gravitational field of the SMBH rips the star apart \cite{1975Natur.254..295H}.  Although subsequent work showed that TDE rates were insufficient to fuel most active galactic nuclei (AGN) emission \cite{1976MNRAS.176..633F,1977ApJ...212..367Y,1999MNRAS.309..447M}, TDEs could still power bright electromagnetic flares lasting several years in galaxies hosting SMBHs at their centers \cite{rees88}.  Several TDE candidates were discovered in the ROSAT all-sky survey \cite{bade96,KomossaBade1999,1999A&A...349L..45K,1999A&A...350L..31G,2000A&A...362L..25G}; see \cite{2020SSRv..216..124V,2020SSRv..216...85S,2020SSRv..216...81A} for recent reviews of the observed optical/UV, X-ray, and radio properties of TDE candidates.

After tidal disruption, about half of the mass of the star remains bound to the SMBH and forms a stream, with each stream element orbiting the SMBH on its own highly eccentric orbit \cite{1982ApJ...262..120L}. Upon returning to pericenter, the leading stream elements experience relativistic apsidal precession, causing their orbits to intersect with those of trailing elements \cite{1994ApJ...422..508K}.  Energy dissipation in the resulting stream self-collisions may produce prompt optical/UV emission \cite{2015ApJ...806..164P,2016ApJ...830..125J} and circularize the stream orbits, promoting the subsequent formation of an accretion disk about the SMBH \cite{shiokawa15,2015ApJ...812L..39D,2016MNRAS.455.2253B,2016MNRAS.461.3760H,bonnerot17,2020MNRAS.492..686L,2020MNRAS.495.1374B,2021SSRv..217...16B}.

It was initially suggested that the timescale for this tidal-stream circularization would be shorter than the fallback time of the most tightly bound tidal debris, implying that the bolometric TDE light curve would trace the mass fallback rate onto the SMBH \cite{rees88}.  Multi-band photometry of the TDE candidate PS1-10jh \cite{gezari12} was well fit by a numerical model \cite{lodato09} predicated on this assumption.
However, observations of two TDE candidates found by the All-Sky Automated Survey for SuperNovae (ASASSN) showed different time evolution in the optical/UV and X-ray light curves, suggesting that both could not simultaneously trace the mass fallback rate.  X-ray variations in the TDE candidate ASASSN-14li were correlated with optical/UV fluctuations with a lag of $32 \pm 4$ days \cite{pasham17}.  One proposed explanation was that the optical/UV emission was produced promptly in the stream self-collisions, while the correlated X-ray emission was produced when the same stream elements returned to pericenter.  Observations of the TDE candidate ASASSN-15oi between 200 and 400 days after its discovery showed that the X-ray emission had increased by an order of magnitude while the UV/optical emission dropped by a factor of 100 \cite{gezari17}.  This delay could be explained by inefficiency in tidal-stream circularization, with the early UV/optical emission generated in the stream self-collisions and the late-time X-ray brightening occurring after the accretion disk had fully assembled.  An alternative model for ASASSN-15oi is that an optically thick outflow reprocesses X-ray emission from the inner disk into early UV/optical emission \cite{2016MNRAS.461..948M}, but these X-rays can reveal themselves at late times once the outflow is fully ionized.  It is unclear how reprocessing could cause the UV/optical emission to lead the X-ray emission as in ASASSN-14li.

Inspired by these observations and previous theoretical work,  we have undertaken a new kinematic study of tidal-stream circularization.  We assume that the tidal stream is one-dimensional and confined to the initial orbital plane of the star, and that the collision is fully inelastic, i.e. the leading edge of the part of the stream that has passed through pericenter (stream I) merges completely into the part of the stream that is returning to pericenter for the first time (stream II).
\citeauthor{bonnerot17}~\cite{bonnerot17} made these same assumptions and also assumed that colliding stream elements had the same mass and orbital energy.  This latter assumption greatly simplified their model of the tidal-stream evolution, but is inconsistent with the time-dependent mass fallback rate.  It necessitated the artificial decomposition of the luminosity of the stream collision into two distinct components: a "stream self-crossing shock luminosity" associated with stream I colliding with itself, and a "tail shock luminosity" associated with the merger of stream II into stream I.  By self-consistently tracking the mass along the lengths of our one-dimensional streams I and II, we are able to preserve the integrity of the single stream-collision point and model the mass ratio of the colliding streams.

We use our new model to predict the efficiency of tidal-stream circularization as a function of SMBH mass and penetration factor.  We hope that it will provide a useful intermediate step between the even more simplified model of \citeauthor{bonnerot17}~\cite{bonnerot17} and vastly more sophisticated general relativistic hydrodynamics simulations such as those presented in \citeauthor{shiokawa15}~\cite{shiokawa15}.  We describe our methodology in much greater detail in Sec.~\ref{S:Method}, present the predictions of our model in Sec.~\ref{S:results}, then briefly discuss their implications in Sec.~\ref{S:disc}.  A short appendix examines how kinematic effects cause our assumption of a one-dimensional stream to break down beyond a certain time in our simulation.

\section{Methodology} \label{S:Method}

\subsection{Initial conditions}

We consider a star of mass $M_\star$ and radius $R_\star$ approaching a non-spinning SMBH of mass $M_\bullet$ on a parabolic orbit with specific angular momentum
\begin{equation}
L = \left( \frac{2GM_\bullet r_t}{\beta} \right)^{1/2} = L_t \beta^{-1/2},
\end{equation}
where $r_t = (M_\bullet/M_\star)^{1/3} R_\star$ is the tidal radius, the penetration factor $\beta = r_t/r_p$ is the ratio of tidal and pericenter radii, and $L_t \equiv (2GM_\bullet r_t)^{1/2}$ is the specfic angular momentum of an orbit with $\beta = 1$.  The extreme mass ratio $q \equiv M_\star/M_\bullet \ll 1$ between the star and SMBH implies a hierarchy between the specific self-binding energy $E_\star \equiv GM_\star/R_\star$ of the star and the specific binding energy
\begin{equation}
E_t \equiv \frac{GM_\bullet R_\star}{r_t^2} = q^{-1/3} E_\star \gg E_\star~.
\end{equation}
of the most-bound tidal-stream element following tidal disruption.

\begin{figure}[t!]
\includegraphics[width=\linewidth]{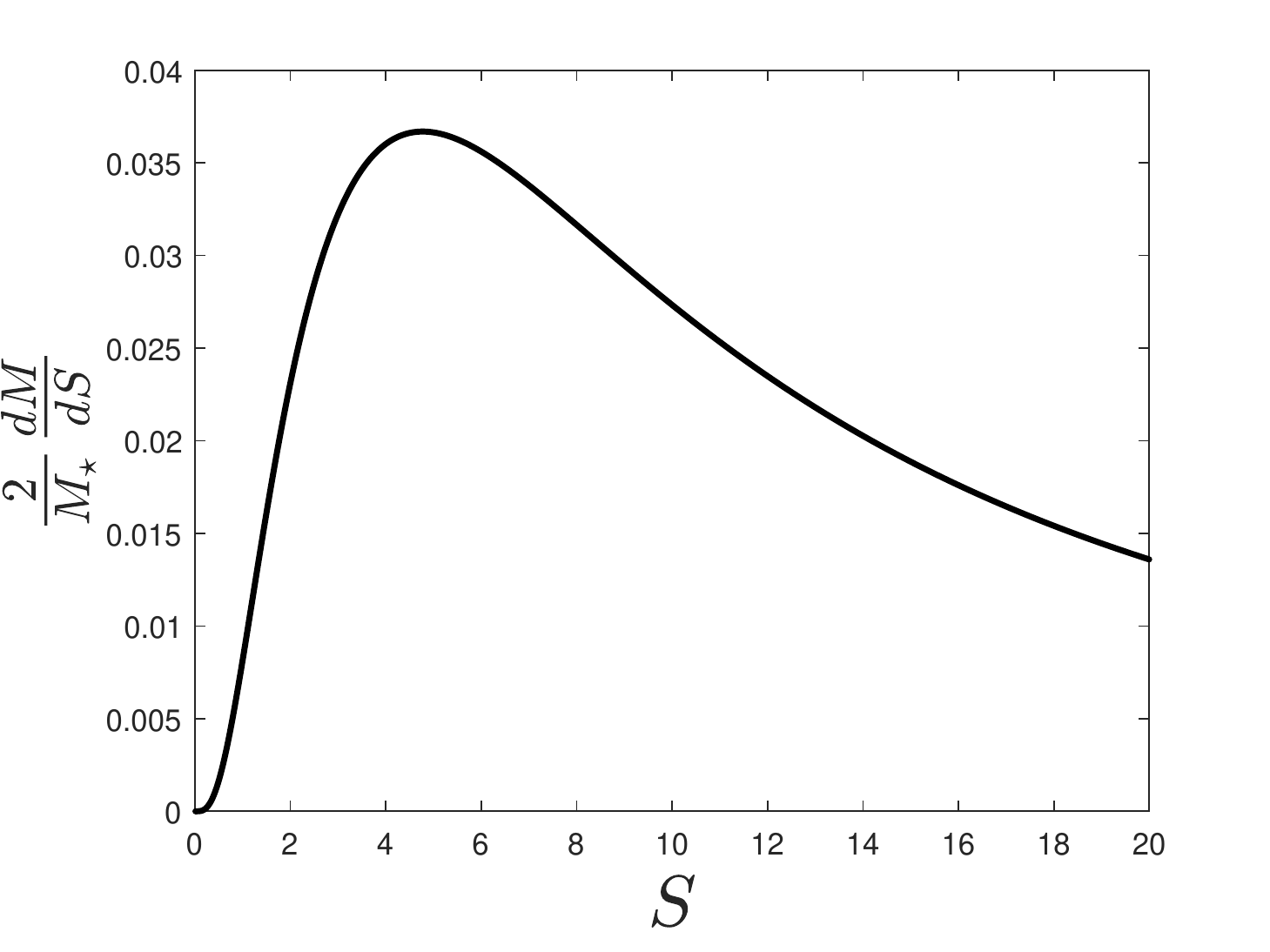}
\caption{The mass per unit dimensionless fallback time $dM/dS$ as a function of dimensionless fallback time $S$ in the freeze-in model of \citeauthor{lodato09}~\cite{lodato09} applied at the tidal radius $r_t$ for a star with polytropic index $n = 3$.}
\label{fig:dm_dS}
\end{figure}

We parameterize the tidal debris distribution using the dimensionless fallback time 
\begin{equation} \label{S definition}
S(E_i) \equiv \frac{\tau(E_i)}{\tau_{\rm mb}} - 1 = \left( \frac{E_i}{E_t} \right)^{-3/2} - 1~,
\end{equation}
where $\tau(E_i)$ is the orbital period of a tidal-debris element of initial specific binding energy $E_i$, $\tau_\star = (R_\star^3/GM_\star)^{1/2}$ is the stellar dynamical timescale, and
\begin{align}
\tau_{\rm mb} &= \tau(E_t) =  \frac{\pi\tau_\star}{(2q)^{1/2}} \notag \\
&= 0.112~{\rm yr} \left( \frac{M_\bullet}{10^6 M_\odot} \right)^{1/2} \left( \frac{M_\star}{M_\odot} \right)^{-1} \left( \frac{R_\star}{R_\odot} \right)^{3/2}
\end{align}
is the initial orbital period of the most tightly bound tidal debris.  Since $E_t > E_i > 0$, Eq.~(\ref{S definition}) implies that $0 < S < \infty$ with $S = 0$ ($S = \infty$) labeling the leading (trailing) edge of the tidal stream.

In terms of this new parameter, we calculate the initial mass distribution $dM/dS$ along the tidal stream using the "freeze-in" model of \citeauthor{lodato09}~\cite{lodato09} in which the specific energy distribution of the tidal debris immediately following disruption is equated to the distribution of gravitational potential for a spherical star.  We adopt a polytropic index $n = 3$ appropriate for a solar-type star and apply the freeze-in model at the tidal radius where disruption occurs \cite{stone13}.  The hierarchy $E_\star \ll E_t$ makes this model an excellent approximation for the extreme mass ratios $q < 10^{-5}$ relevant to the tidal disruption of main-sequence stars by SMBHs.

This initial mass distribution $dM/dS$ is shown in Fig.~\ref{fig:dm_dS}.  It is independent of $M_\bullet$ and $\beta$ for our dimensionless parameters, and the total mass $M_\star/2$ of the bound tidal stream implies that the area under the curve is unity for the normalization adopted in this figure.  In the freeze-in model, each stream element $S$ corresponds to a slice through the star perpendicular to the line of separation between the star and SMBH a distance $R(S) = R_\star (1 + S)^{-2/3}$ from the center of the star.  This implies that $dM/dS = 0$ at the leading edge of the stream ($S = 0$), since this corresponds to a slice of cross section $\pi[R_\star^2 - R^2(S)] = 0$.  The distribution reaches a maximum value of $(2/M_\star) dM/dS = 0.0367$ at $S_{\rm max} = 4.78$, and half of its mass is found at $S < S_{1/2} = 22$.  Along the trailing edge of the stream, it approaches the limit
\begin{align} \label{E:dMdSLate}
\lim_{S \to \infty} \frac{2}{M_\star} \frac{dM}{dS} &= \frac{4R_\star}{3M_\star} \frac{dM_\perp}{dR}(0) S^{-5/3} \approx 2.95 S^{-5/3}~,
\end{align}
where $dM_\perp/dR$ is the mass per unit thickness of a slice of the star passing a minimum distance of $R$ from its center.  The scaling $\propto S^{-5/3}$ in Eq.~(\ref{E:dMdSLate}) is consistent with the late-time scaling $\propto t^{-5/3}$ of the mass fallback rate \cite{rees88,1989IAUS..136..543P}.

We choose the origin of our coordinate system to be the location of the SMBH, the $xy$ plane to coincide with the orbital plane of the star, and the $x$ axis to be along the line of separation between the SMBH and star at tidal disruption (when $r = r_t$).  In this coordinate system, the initial position, specific binding energy, and specific orbital angular momentum of the tidal-stream elements are given by
\begin{subequations} \label{E:rEL}
\begin{align}
\frac{r_i}{r_t} &= 1 - \frac{R(S)}{r_t} = 1 - q^{1/3}\frac{E_i}{E_t} \notag \\
&= 1 - q^{1/3}(1 + S)^{-2/3}, \\
\frac{E_i}{E_t} &= \frac{GM_\bullet}{E_t}\left( \frac{1}{r} - \frac{1}{r_t} \right) \approx (1 + S)^{-2/3}, \label{E:Ei} \\
\frac{L}{L_t} &= \beta^{-1/2} \left[ 1 - \frac{R(S)}{r_t} \right]  \approx \beta^{-1/2},
\end{align}
\end{subequations}
where $R(S)$ is the distance of stream element $S$ from the center of the star at tidal disruption and the approximations are to lowest order in $R(S)/r_t \leq q^{1/3} \ll 1$.  Eqs.~(\ref{E:rEL}) can be used to calculate the initial semi-major axis, eccentricity, true anomaly, and argument of pericenter
\begin{subequations} \label{E:iKep}
\begin{align}
\frac{a_i}{r_t} &= \frac{GM_\bullet}{2E_ir_t} = \frac{q^{-1/3}}{2}(1 + S)^{2/3}, \label{E:ai} \\
1 - e_i^2 &= \frac{L^2}{GM_\bullet a_i} = \frac{4q^{1/3}}{\beta} (1 + S)^{-2/3}, \label{E:ei} \\
\cos f_i &= \frac{1}{e_i} \left[\frac{a_i(1 - e_i^2)}{r} - 1 \right] \approx \frac{2}{\beta} - 1~, \label{E:fi} \\
\omega_i &= -f_i~, \label{E:omi}
\end{align}
\end{subequations}
where $f_i$ ($\omega_i$) is negative (positive) for $\beta > 1$.

\subsection{Ballistic evolution} \label{SS:BE}

We assume that the tidal-stream elements orbit the SMBH ballistically between collisions.  This ballistic evolution can be calculated most easily in terms of the eccentric anomaly $\mathcal{E}$ given by
\begin{equation} \label{E:EAdef}
\tan\frac{\mathcal{E}}{2} = \left( \frac{1 - e}{1 + e} \right)^{1/2} \tan\frac{f}{2}~.
\end{equation}
Inserting Eqs.~(\ref{E:ei}) and (\ref{E:fi}) into Eq.~(\ref{E:EAdef}) yields the initial eccentric anomaly $\mathcal{E}_i(S) < 0$ and thus the time
\begin{equation} \label{E:t0i}
\frac{t_0(S)}{\tau_{\rm mb}} = \frac{1 + S}{2\pi}(e_i\sin\mathcal{E}_i - \mathcal{E}_i)
\end{equation}
at which stream element $S$ first passes through pericenter.  This equation indicates that $t_0 = 0$ when tidal disruption occurs at pericenter ($\beta = 1$), while $t_0 \ll \tau_{\rm mb}$ for all reasonable values of $\beta$ given that the initial apocenter satisfies $r_a \approx 2a_i \gg r_t$ according to Eq.~(\ref{E:ai}).

The subsequent time evolution of the eccentric anomaly is then found by solving the transcendental equation
\begin{equation} \label{E:EAt}
\mathcal{E}(S, t) - e(S)\sin\mathcal{E}(S,t) = \frac{[2E(S)]^{3/2}}{GM_\bullet}[t - t_0(S)]~.
\end{equation}
Given that relativistic apsidal precession is strongly peaked near pericenter on the highly eccentric orbits of the tidal-stream elements, we model it as an instantaneous increment in the argument of pericenter at each pericenter passage ($\mathcal{E} = 2\pi n$) by an amount
\begin{align} \label{E:PrecessionAngle}
\Delta\omega &= 6\pi\left(\frac{GM_\bullet}{Lc}\right)^2 \notag \\
&= 0.2\beta \left(\frac{M_\bullet}{10^6M_\odot}\right)^{2/3} \left(\frac{M_\star}{M_\odot}\right)^{1/3} \left(\frac{R_\star}{R_\odot}\right)^{-1} 
\end{align}
given by the first post-Newtonian correction to the equation of motion \cite{2013degn.book.....M}.  The position of stream element $S$ according to this ballistic evolution is given by
\begin{subequations} \label{E:PolCoord}
\begin{align}
\frac{r(S, t)}{r_t} &= \frac{a(1 - e^2)}{r_t(1 + e \cos f)} = \frac{2}{\beta[1 + e(S)\cos f(S, t)]}~, \\
\phi(S, t) &= f(S, t) + \omega(S)~,
\end{align}
\end{subequations}
where $f(S, t)$ is found from Eqs.~(\ref{E:EAdef}) and (\ref{E:EAt}), and its velocity is given by
\begin{subequations} \label{E:PolVel}
\begin{align}
\frac{v_r(S, t)}{v_t} &= \frac{GM_\bullet e\sin f}{Lv_t} = \frac{\beta^{1/2}}{2}  e(S)\sin f(S, t)~, \label{E:RV} \\
\frac{v_\phi(S, t)}{v_t} &= \frac{L}{rv_t} = \frac{\beta^{1/2}}{2} [1 + e(S)\cos f(S, t)]~, \label{E:TV}
\end{align}
\end{subequations}
where
\begin{equation} \label{E:vt}
v_t \equiv \frac{L_t}{r_t} = q^{-1/3} (2E_\star)^{1/2}. 
\end{equation}

\subsection{Initial stream self-intersection} \label{SS:SI}

\begin{figure}[t!]
\includegraphics[width=\linewidth]{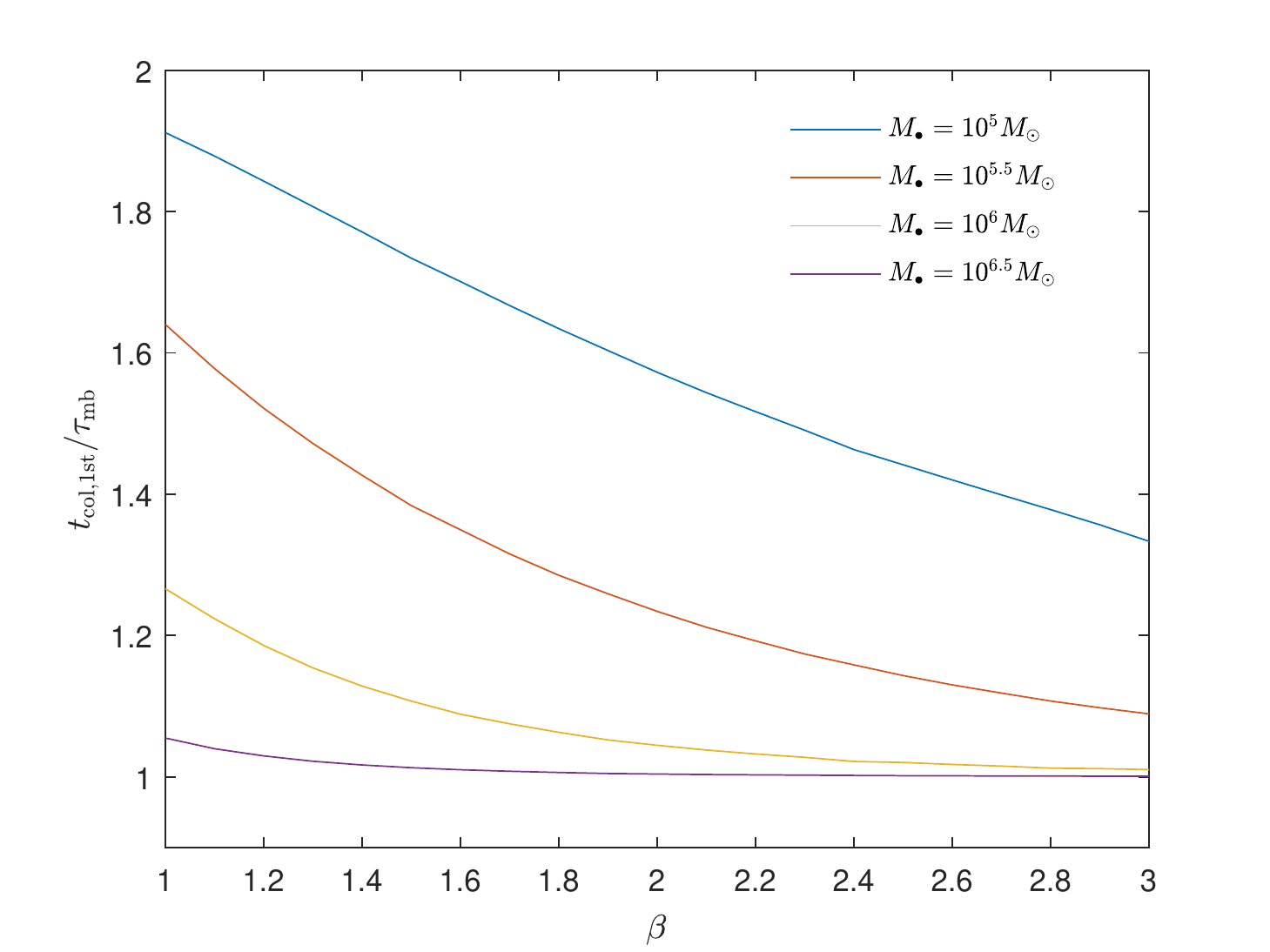}
\caption{Time of the first stream collision in units of the period of the most bound stream element as a function of penetration factor $\beta$ for SMBH masses $M_\bullet$ of $10^5M_\odot$, $10^{5.5}M_\odot$, $10^6M_\odot$, and $10^{6.5}M_\odot$.}
\label{fig:t_col}
\end{figure}

Evolving the positions of the stream elements according to Eq.~(\ref{E:PolCoord}), we find that the leading element ($S = 0$) first collides with a trailing stream element at a time $t_{\rm col,1st}$ shown as a function of penetration factor $\beta$ for different SMBH masses $M_\bullet$ in Fig.~\ref{fig:t_col}.  For small $\beta$ and $M_\bullet$, apsidal precession is negligible according to Eq.~(\ref{E:PrecessionAngle}).  In this limit, the first collision occurs when the leading element returns to the tidal radius for the second time and laps the element $S = 1$ that is returning for the first time ($t_{\rm col,1st} \to 2\tau_{\rm mb}$).  In the opposite limit of large $\beta$ and $M_\bullet$, strong apsidal precession causes the leading element to collide almost immediately after its first return to pericenter ($t_{\rm col,1st} \to t_0(0) + \tau_{\rm mb} \approx  \tau_{\rm mb}$).

\begin{figure*}[t!]
\begin{subfigure}[t]{0.49\textwidth}
\includegraphics[width=\linewidth]{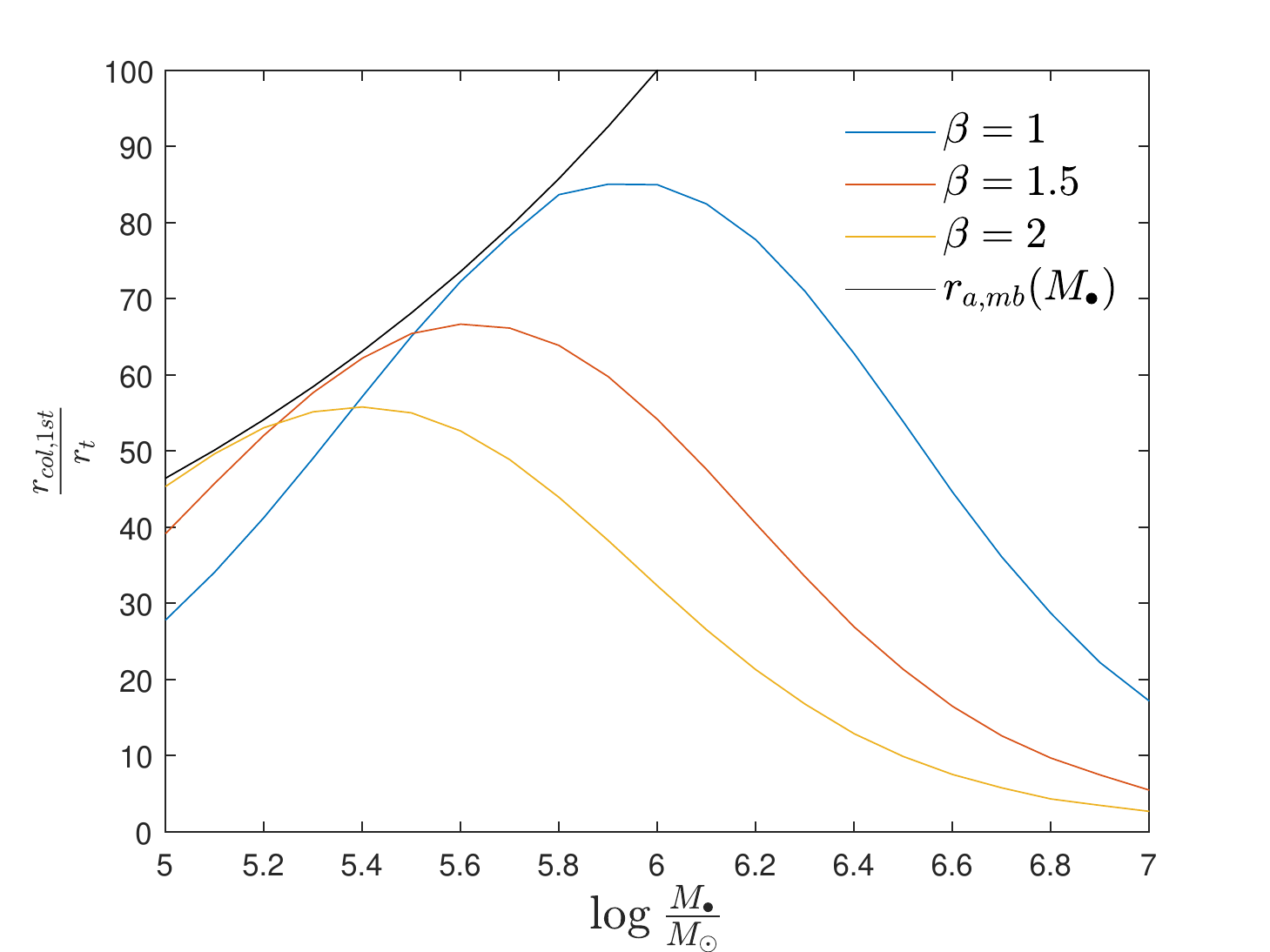}
\end{subfigure}
\begin{subfigure}[t]{0.49\textwidth}
\includegraphics[width=\linewidth]{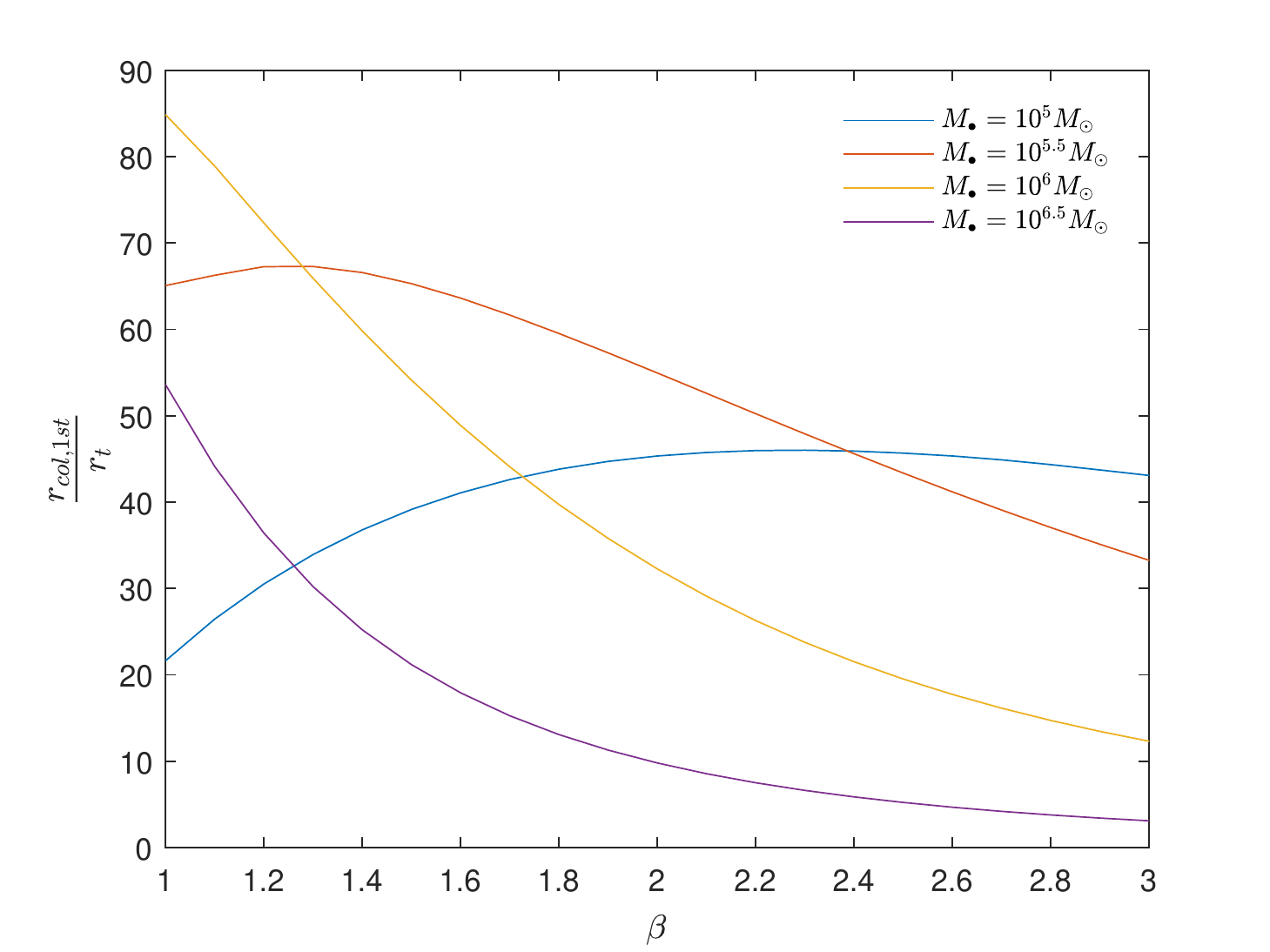}
\end{subfigure}
\caption{Distance $r_{\rm col}$ of the initial collision point from the SMBH in units of the tidal radius $r_t$.  The left panel shows this distance as a function of SMBH mass $M_\bullet$ for penetration factors $\beta$ of 1, 1.5, and 2. The black curve shows the apocenter distance $r_{a, {\rm mb}}/r_t \approx q^{-1/3}$ of the most bound stream element.  The right panel shows $r_{\rm col}$ as a function of $\beta$ for $M_\bullet = 10^5M_\odot$, $10^{5.5}M_\odot$, $10^6M_\odot$, and $10^{6.5}M_\odot$.}
\label{fig:r_col}
\end{figure*}

The distance $r_{\rm col, 1st}$ of this initial collision point from the SMBH as a function of SMBH mass and penetration factor is shown in Fig.~\ref{fig:r_col}.  This distance would approach the tidal radius for negligible apsidal precession, but this limit is not achieved even for $M_\bullet = 10^5 M_\odot$, $\beta = 1$ for which $\Delta\omega \approx 2.5^\circ$ according to Eq.~(\ref{E:PrecessionAngle}).  As $M_\bullet$ increases in the left panel, apsidal precession increases and the initial collision point is located further from the tidal radius.  The black curve shows the initial apocenter $r_{a,{\rm mb}} \approx q^{-1/3}r_t$ of the most tightly bound stream element; the initial collision occurs at apocenter ($t_{\rm col,1st} = t_0(0) + 1.5\tau_{\rm mb}$) for values of $M_\bullet$ at which this curve is tangent to the colored curves showing $r_{\rm col, 1st}$ for different values of $\beta$.  As $M_\bullet$ increases beyond this point, the initial collision occurs closer to pericenter.  Although $r_{\rm col, 1st}$ continues to increase because the initial semi-major axis $a_i \propto M_\bullet^{1/3}$ according to Eq.~(\ref{E:ai}), it eventually reaches a maximum and turns over as $r_{\rm col, 1st} \to r_t/\beta$ in the limit of large apsidal precession.  This maximum occurs at smaller values of $M_\bullet$ for larger penetration factors because $\Delta\omega \propto \beta$ according to Eq.~(\ref{E:PrecessionAngle}).

The right panel of Fig.~\ref{fig:r_col} shows $r_{\rm col, 1st}$ as a function of penetration factor $\beta$ for four different SMBH masses $M_\bullet$.  The two smaller masses, $M_\bullet = 10^5 M_\odot$ and $10^{5.5} M_\odot$, are below the value of $M_\bullet$ of the tangent point between the apocenter curve $r_{a,{\rm mb}}(M_\bullet)$ and the $r_{\rm col, 1st}(M_\bullet)$ curve for $\beta = 1$ in the left panel.  This implies that $r_{\rm col, 1st}(\beta)$ will increase for these SMBH masses until the initial collision occurs at apocenter, then decreases as $r_{\rm col, 1st} \to r_t/\beta$ in the limit of large apsidal precession.  For the two larger masses, $M_\bullet = 10^6 M_\odot$ and $10^{6.5} M_\odot$, the initial collision happens before the second passage through apocenter ($t_{\rm col,1st} < 1.5\tau_{\rm mb}$ in Fig.~\ref{fig:t_col}) even for $\beta = 1$.  The distance of the initial collision point from the SMBH is therefore a monotonically decreasing function of $\beta$ for these masses.

\subsection{Stream circularization} \label{SS:KS}

\begin{figure*}[t!]
\begin{subfigure}[b]{0.49\textwidth}
\includegraphics[width=\linewidth]{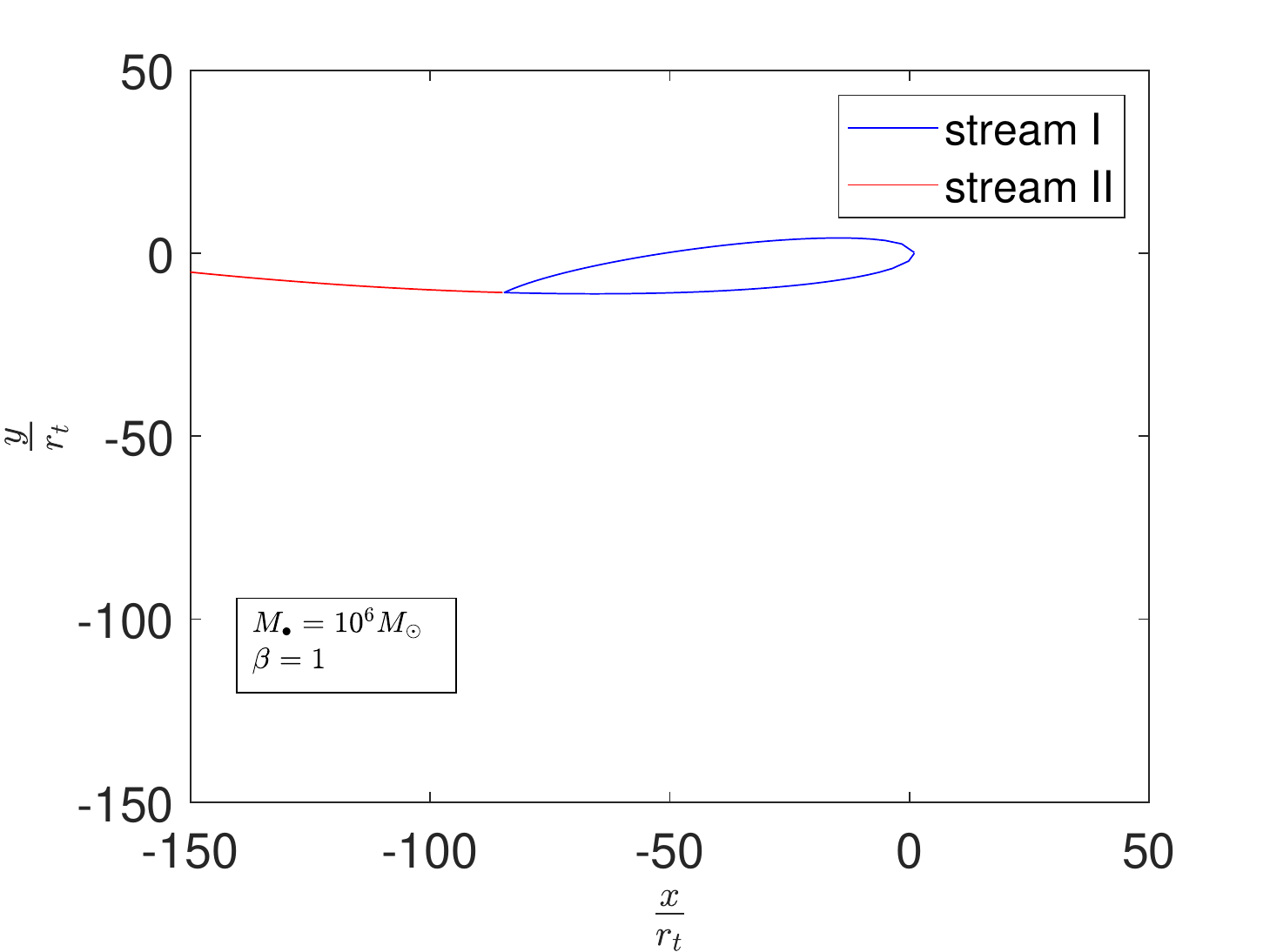}
\end{subfigure}
\begin{subfigure}[b]{0.49\textwidth}
\includegraphics[width=\linewidth]{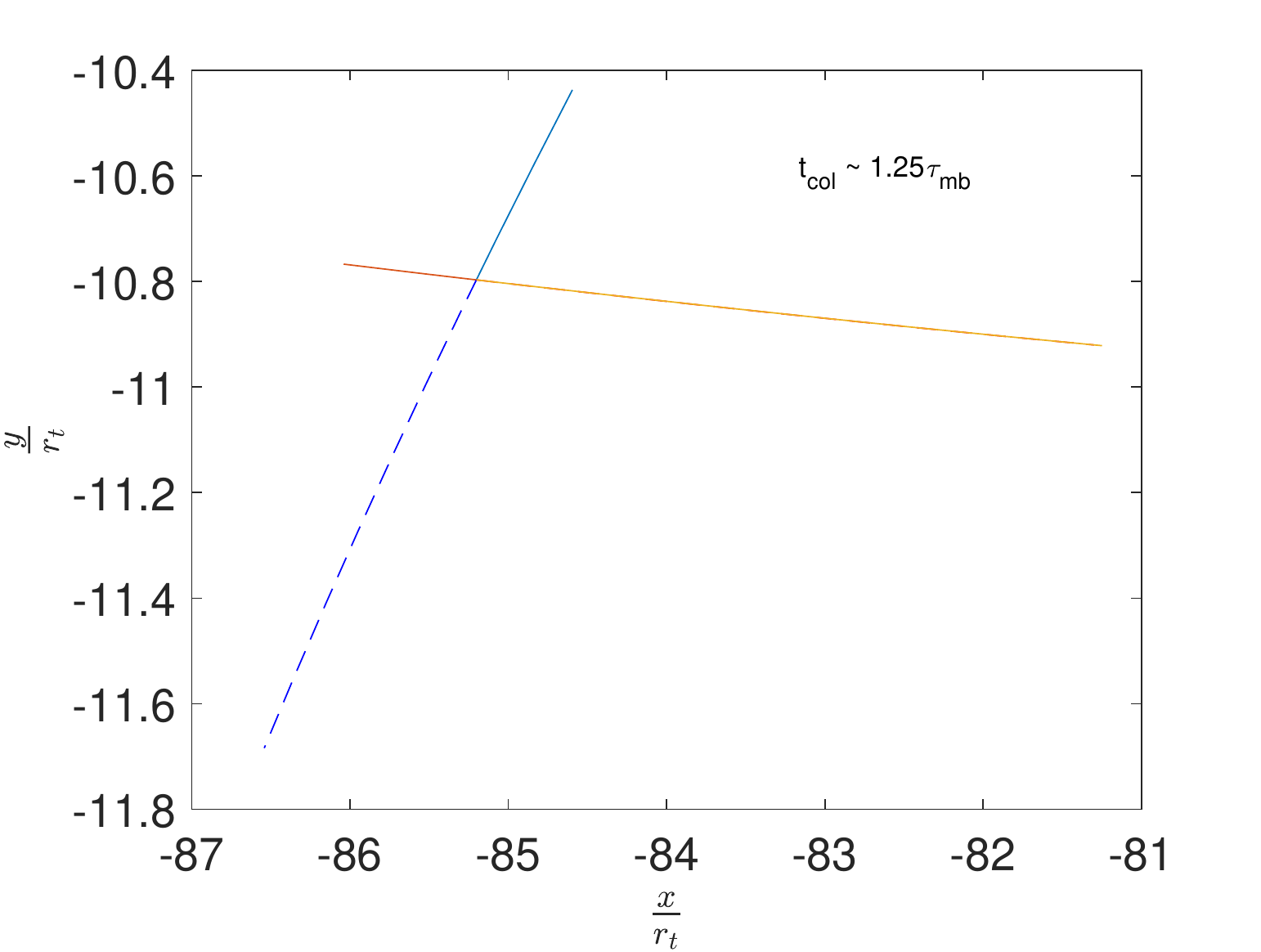}
\end{subfigure}
\begin{subfigure}[b]{0.49\textwidth}
\includegraphics[width=\linewidth]{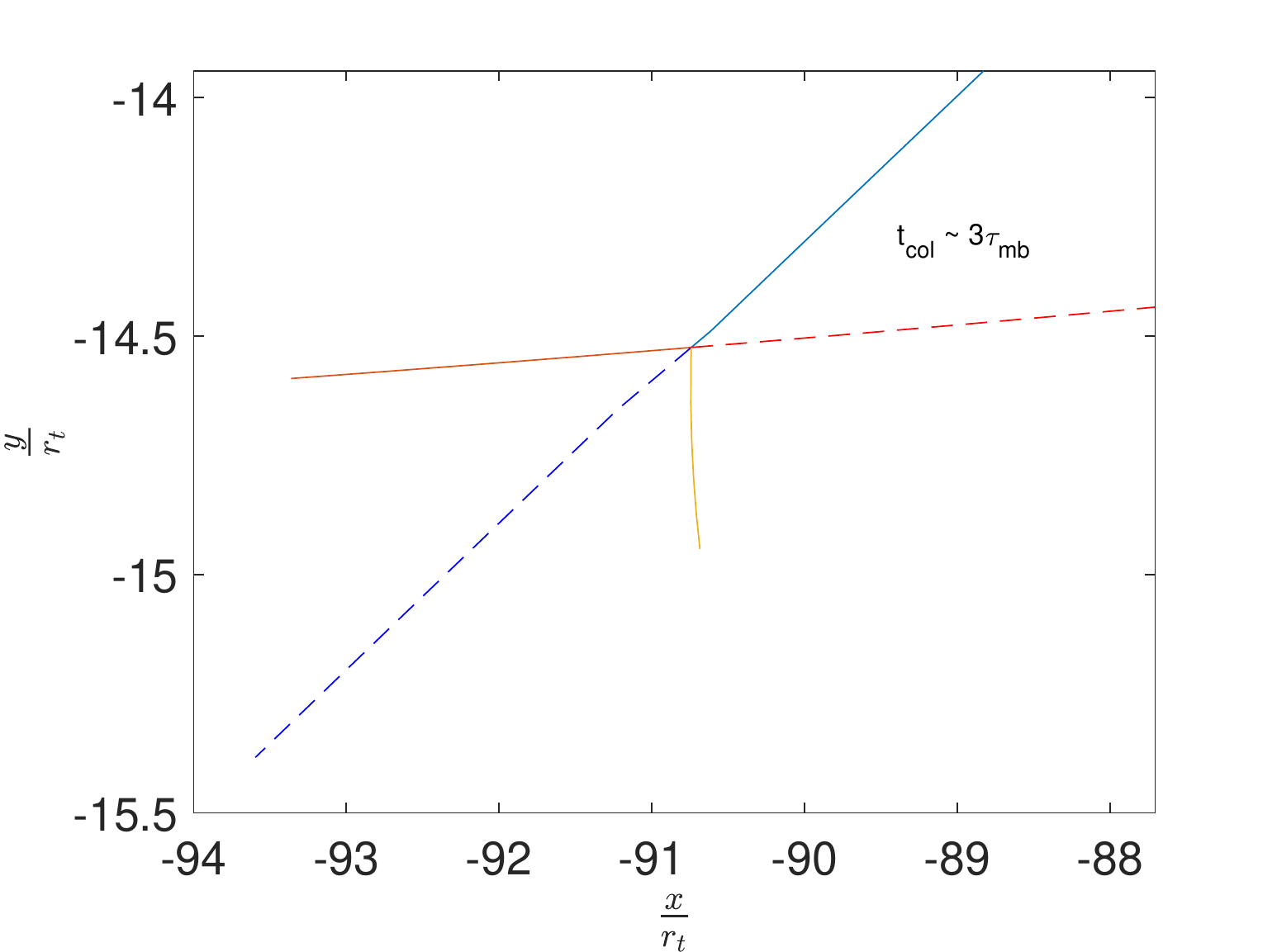}
\end{subfigure}
\begin{subfigure}[b]{0.49\textwidth}
\includegraphics[width=\linewidth]{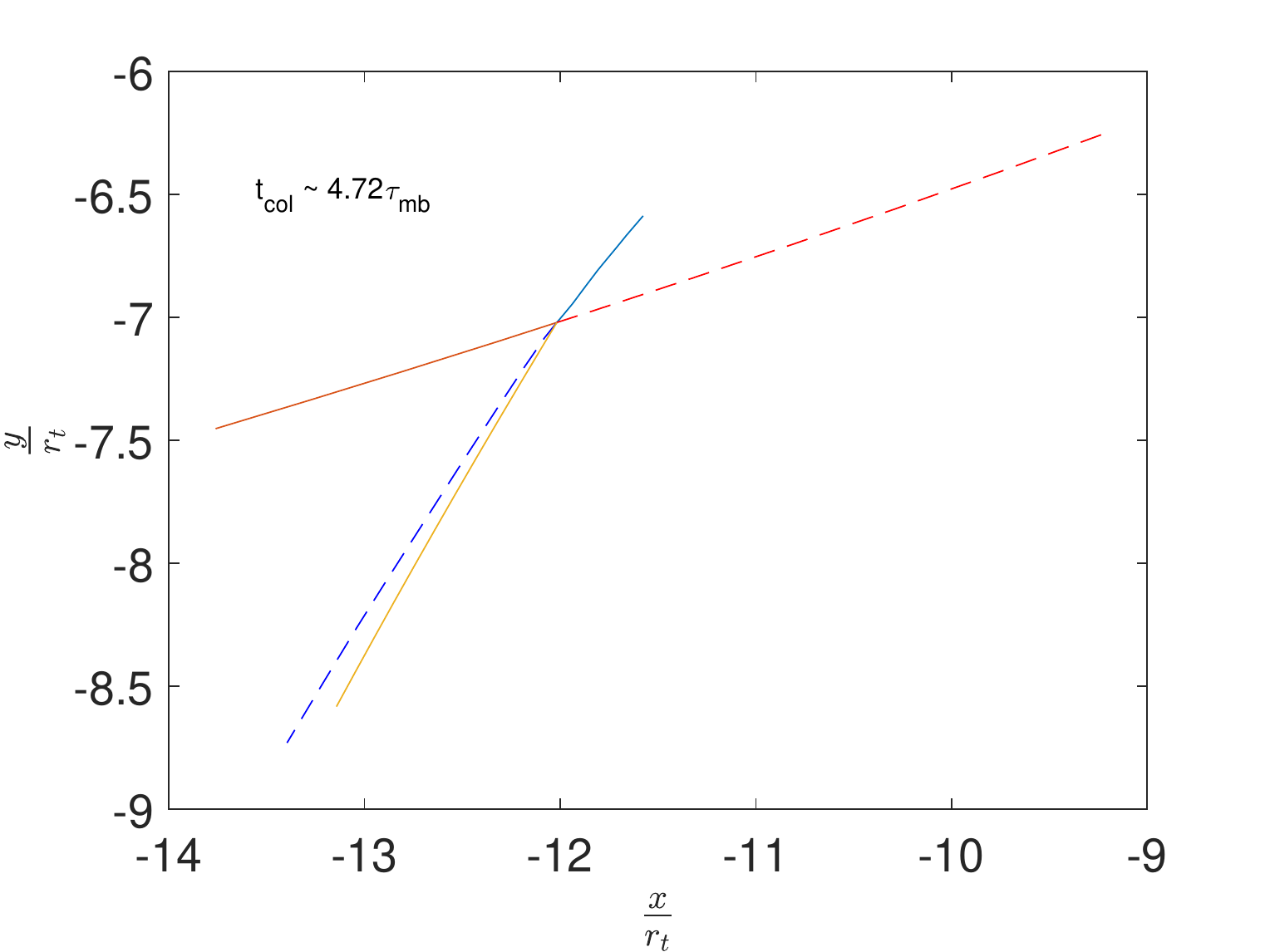}
\end{subfigure}
\caption{Snapshots of the tidal stream after its first self-intersection for $M_\bullet = 10^6 M_\odot$ and $\beta = 1$.  The top left panel is at the time of the first self-intersection; stream I (II) will reach the collision point after (before) returning to the tidal radius.  The remaining panels show the collision point at times $t_{\rm col}/\tau_{\rm mb} = 1.25$, 3, and 4.72 after tidal disruption.  Stream I (II) is shown before the collision by the solid blue (red) line and extrapolated after the collision by the dashed blue (red) line.  Stream III is formed in the collision and shown by the solid yellow line.}
\label{fig: stream reprocessing}
\end{figure*}

After the initial stream self-intersection described in the previous section, we assume that streams I and II merge inelastically into a new stream III.  If $S_1(t)$ and $S_2(t)$ are the elements of streams I and II reaching the collision point at time $t$, conservation of mass implies that the mass distribution of the newly formed stream element $S_3(t)$ of stream III will be given by
\begin{equation} \label{E:MassCon}
\frac{dM}{dS}(S_3) =  [1 + Q(t)] \frac{dM}{dS}(S_2)~, 
\end{equation}
where
\begin{equation} \label{E:Q}
Q(t) \equiv \frac{dM}{dS}(S_1) \left[ \frac{dM}{dS}(S_2) \right]^{-1} \frac{dS_1/dt}{dS_2/dt} 
\end{equation}
is the ratio of the mass flux of stream I to II at the collision point and the new element of stream III inherits the label of its parent from stream II ($S_3 = S_2$).  This inelastic collision preserves specific orbital angular momentum (and thus the azimuthal velocity $v_\phi$) but not specific energy.  The radial velocity $v_r$ of the new element of stream III is given by conservation of linear momentum:
\begin{equation} \label{E:MomCon}
v_r(S_3) = \frac{v_r(S_2) + Q(t)v_r(S_1)}{1 + Q(t)}~.
\end{equation}
Using the position of the collision point and the velocity of the new element of stream III, we can calculate its semi-major axis $a(S_3)$, eccentricity $e(S_3)$, argument of pericenter $\omega(S_3)$, and eccentric anomaly $\mathcal{E}(S_3)$ using the standard definitions of the Keplerian orbital elements.  We then evolve this element balistically as described in Sec.~\ref{SS:BE} until it collides with a new element of stream II when it returns to the collision point.

Four snapshots of this evolution for the tidal disruption of a solar-type star with penetration factor $\beta = 1$ by a SMBH of mass $10^6 M_\odot$ are shown in Fig.~\ref{fig: stream reprocessing}.  The top left panel shows the entirety of stream I at the time of the initial stream self-intersection.  This intersection occurs at $t_{\rm col,1st} \approx 1.25\tau_{\rm mb}$ at a collision point a distance $r_{\rm col,1st} \approx 85r_t$ from the SMBH consistent with Figs.~\ref{fig:t_col} and \ref{fig:r_col}.  The top right panel shows a close-up view of the collision point shortly after this initial self-intersection.  A key feature of this panel is that the newly formed stream III is nearly parallel to the extrapolation of stream II through the collision point.  This can be understood by recognizing that at the time $t_{\rm col,1st}$ of the initial stream self-intersection, $S_1 = 0$ and therefore $dM/dS(S_1) = 0$ as shown in Fig.~\ref{fig:dm_dS}.  This requires $Q(t_{\rm col,1st}) = 0$ according to Eq.~(\ref{E:Q}) and therefore $v_r(S_3) = v_r(S_2)$ according to Eq.~(\ref{E:MomCon}).  This feature is crucial because it implies that the leading edge of stream III smoothly connects to the trailing edge of stream I.  The topology of the tidal stream therefore always consists of a single closed loop of stream I/III with the infalling stream II proving a tail at the collision point as in the top left panel of Fig.~\ref{fig: stream reprocessing}. There is no need to assume that stream I self-intersects at a different location providing a "stream self-crossing shock luminosity" distinct from the "tail shock luminosity" as in \citeauthor{bonnerot17}~\cite{bonnerot17}.

The bottom two panels in Fig.~\ref{fig: stream reprocessing} show the collision point at later times.  As $S_1(t)$ increases, $dM/dS(S_1)$ and thus $Q(t)$ increase rapidly.  They increase even more rapidly than shown in Fig.~\ref{fig:dm_dS} once the leading element $S_3(t_{\rm col,1st})$ of stream III returns to the collision point, but they remain continuous because $Q(t_{\rm col,1st}) = 0$.  The angle between stream III and stream II increases with time as $Q(t)$ increases; stream III is nearly parallel to stream I ($v_r(S_3) \approx v_r(S_1)$) at the final time $t_{\rm col} = 4.72\tau_{\rm mb}$ depicted in the bottom right panel of Fig.~\ref{fig: stream reprocessing}.

\subsection{Extrapolation} \label{SS:extrap}

The final time $t_{\rm col} = 4.72\tau_{\rm mb}$ shown in Fig.~\ref{fig: stream reprocessing} was not set by choice, but because our model breaks down shortly after this time.  This breakdown, explored in detail in the Appendix, is caused by the rapid increase in $Q(t)$ and thus the energy dissipated at the collision point.  Trailing elements of stream III develop higher binding energies and thus shorter orbital periods than leading elements.  When this effect becomes large enough, the trailing elements are able to overtake the leading elements before they have passed through the collision point.  A kink is formed in the tidal stream as shown in Fig.~\ref{F:kink}, breaking the "loop + tail" topology essential to our model.  A hydrodynamics simulation could address this feature by forming a shock, and we speculate that this effect could contribute to the delayed formation of shock 2b in the hydrodynamics simulation of \citeauthor{shiokawa15}~\cite{shiokawa15}.  As our purely kinematic model does not include shocks, we have developed an even more approximate model to extrapolate beyond the time of this breakdown.

This new approximate model extrapolates the true anomaly $f_2(S_2)$ of stream II at the collision point by fitting the function
\begin{equation} \label{E:TA2fit}
\cos f_2(S_2) = \frac{1}{2}\tanh[k(S_2 - C)] - \frac{1}{2}  
\end{equation}
to our previous kinematic simulations, where $k$ and $C$ are fitting parameters.  We chose this function because of the limits of the hyperbolic tangent: 
\begin{equation}
\lim_{x \to \mp\infty} \tanh x = \mp 1~.
\end{equation}
This implies that $\cos f_2 \to -1$ ($f_2 \to -\pi$) at early times consistent with $r_{\rm col,1st} \gg r_t$ as shown in Fig.~\ref{fig:r_col}.  At late times, the orbit of stream II will approach a parabola of specific angular momentum $L = L_t\beta^{-1/2}$, while that of stream I will approach a circle of radius
\begin{equation} \label{E:rc}
r_c = \frac{L^2}{GM_\bullet} = \frac{2r_t}{\beta}
\end{equation}
with the same specific angular momentum.  These two orbits intersect at $f_2 = -\pi/2$ ($\cos f_2 = 0$) consistent with Eq.~(\ref{E:TA2fit}) in the limit $S_2 \to \infty$.

Eq.~(\ref{E:TA2fit}) also allows us to calculate the time $t(S_2)$ when element $S_2$ arrives at the collision point.  We use Eq.~(\ref{E:EAdef}) to calculate the eccentric anomaly from the true anomaly, then Eq.~(\ref{E:EAt}) to determine the time of the collision.

We also need to extrapolate from our kinematic simulations to determine the element $S_1(S_2)$ of stream I that collides with element $S_2$ of stream II.  We fit the function
\begin{equation} \label{E:S1S2fit}
S_1(S_2) = S_2 - \frac{\tau_c}{\tau_{\rm mb}} - A\exp(-B S_2)~,
\end{equation}
where 
\begin{equation}
\tau_c = 2\pi\left( \frac{r_c^3}{GM_\bullet} \right)^{1/2} = 8\left( \frac{q}{\beta^3} \right)^{1/2} \tau_{\rm mb}
\end{equation}
is the orbital period at the circularization radius $r_c$ and $A$ and $B$ are fitting parameters.  This function was chosen to satisfy the limit $S_1 \to S_2 - \tau_c/\tau_{\rm mb}$ as $S_2 \to \infty$ because it will take a time $\tau_c$ for the newly formed element $S_1$ of stream III to travel around the loop and return to the collision point as part of stream I.

Once we have fit Eq.~(\ref{E:S1S2fit}) to our kinematic simulations, we can calculate $dS_1/dS_2$ and use conservation of mass (\ref{E:MassCon}) with $Q(t)$ replaced by
\begin{equation} \label{E:Qex}
Q(S_2) \equiv \frac{dM}{dS}(S_1) \left[ \frac{dM}{dS}(S_2) \right]^{-1} \frac{dS_1}{dS_2} 
\end{equation}
to evolve the mass distribution of the tidal stream beyond the end of our kinematic model.  We can use our fit (\ref{E:TA2fit}) and the requirement that $r_1 = r_2$ at the collision point to find the true anomaly of stream I
\begin{equation} \label{E:TA1}
\cos f_1(S_1) = \frac{e(S_2)}{e(S_1)} \cos f_2(S_2)~,
\end{equation}
then use Eq.~(\ref{E:RV}) to determine the radial velocities of streams I and II and conservation of linear momentum (\ref{E:MomCon}) to determine the radial velocity of stream III.  This implies that the specific binding energy of stream III is given by
\begin{align}
E(S_3) &= \frac{GM_\bullet}{r_{\rm col}} - \frac{1}{2}[v_r^2(S_3) + v_\phi^2(S_3)] \notag \\
&= \frac{E(S_2) + QE(S_1)}{1 + Q} + \frac{Q}{2} \left[ \frac{v_r(S_1) - v_r(S_2)}{1 + Q} \right]^2.
\end{align}
This equation allows us to calculate the bolometric luminosity released at the collision point
\begin{align} \label{E:lum}
L(S_2) &= [(1 + Q)E(S_3) - E(S_2) - QE(S_1)] \frac{dM/dS}{dt/dS_2}(S_2) \notag \\
&= \frac{Q[v_r(S_1) - v_r(S_2)]^2}{2(1 + Q)} \frac{dM/dS}{dt/dS_2}(S_2)~.
\end{align}

\subsection{Late-time limit}

In the limit of late times:
\begin{enumerate}

\item the mass flux ratio goes to infinity ($Q \to \infty$),

\item stream I circularizes [$v_r(S_1) \to 0$],

\item stream II is on a parabolic orbit at the circularization radius [$v_r^2(S_2) \to 2GM_\bullet/r_c - v_\phi^2 = \beta E_t/2q^{1/3}$],

\item stream element $S_2$ collides after an orbital period [$S_2(t) \to t/\tau_{\rm mb}$, $dt/dS_2 \to \tau_{\rm mb}$],

\item the dimensionless mass fallback rate approaches the limit of Eq.~(\ref{E:dMdSLate}) [$dM/dS_2 \to 1.475M_\star (t/\tau_{\rm mb})^{-5/3}$].

\end{enumerate}
If we insert the first four of these conditions into Eq.~(\ref{E:lum}), the circularized bolometric luminosity at late times is
\begin{align}
L_\infty(t) &= \frac{\beta}{4q^{1/3}} \frac{E_t}{\tau_{\rm mb}} \frac{dM}{dS_2}[S_2(t)] , \label{E:lumlate} \\
\frac{L_\infty}{L_{\rm Edd}} &= 21.3\beta \left( \frac{M_\bullet}{10^6 M_\odot} \right)^{-5/6} \left\{ \frac{1}{M_\star} \frac{dM}{dS_2}[S_2(t)] \right\} \notag \\
& \quad \times \left( \frac{M_\star}{M_\odot} \right)^{7/3} \left( \frac{R_\star}{R_\odot} \right)^{-5/2}, \label{E:lumlatenorm}
\end{align}
where Eq.~(\ref{E:lumlatenorm}) has been normalized to the Eddington luminosity $L_{\rm Edd} = 4\pi GM_\bullet m_p c/\sigma_t$.  Inserting the fifth condition into Eq.~(\ref{E:lumlatenorm}) yields
\begin{align}
\frac{L_\infty}{L_{\rm Edd}} &= 31.4\beta \left( \frac{M_\bullet}{10^6 M_\odot} \right)^{-5/6} \left( \frac{t}{\tau_{\rm mb}} \right)^{-5/3} \notag \\
& \quad \times \left( \frac{M_\star}{M_\odot} \right)^{7/3} \left( \frac{R_\star}{R_\odot} \right)^{-5/2}. \label{E:lumlatenormt}
\end{align}
This limit shows that the stream self-intersection can indeed power an Eddington-luminosity electromagnetic transient during a TDE provided that the circularization time $t_c$ after which it applies is not too much greater than the minimum fallback time $\tau_{\rm mb}$.

\section{Results} \label{S:results}

We now explore the predictions of our model for the efficiency of tidal-stream circularization and the accompanying bolometric light curve of energy dissipated at the collision point.

\begin{figure}[t!]
\includegraphics[width=\linewidth]{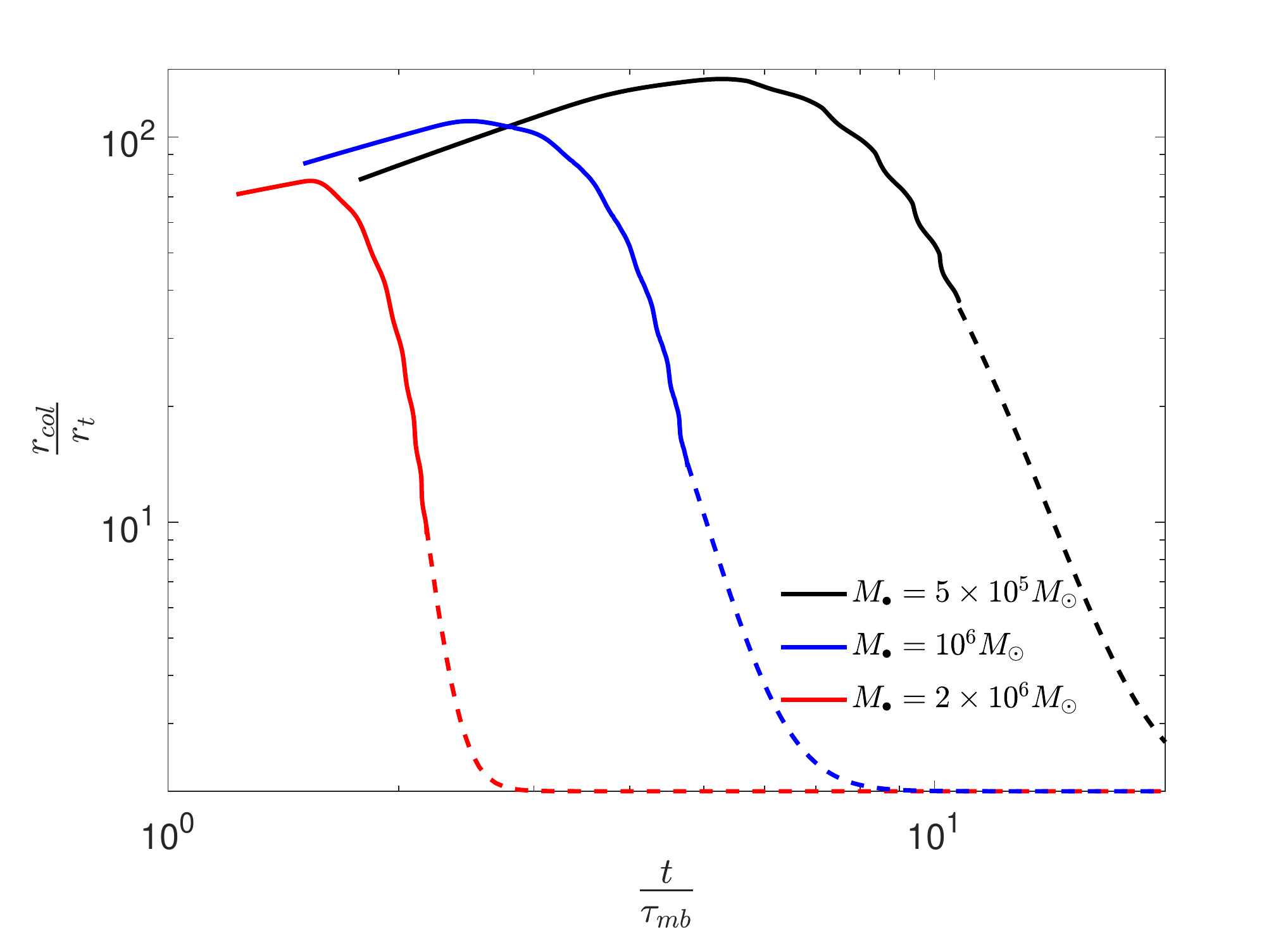}
\caption{Distance $r_{\rm col}$ of the stream collision point from the SMBH as a function of time $t$ after tidal disruption for TDEs with penetration factor $\beta = 1$.  The black, blue, and red curves correspond to SMBH masses of $M_\bullet/M_\odot = 5 \times 10^5$, $10^6$, and $2 \times 10^6$ respectively.  The solid (dashed) portions of each curve show the results of our kinematic simulations and late-time extrapolations.}
\label{fig:r_colVtime}
\end{figure}

The distance $r_{\rm col}(t)$ of the collision point from the SMBH evolves with time as shown in Fig.~\ref{fig:r_colVtime}.   The initial self-intersection occurs at $r_{\rm col,1st} \lesssim r_{a,{\rm mb}} \approx q^{-1/3} r_t$.  At first, the distance $r_{\rm col}(t)$ increases with time, because the trailing elements $S$ of stream I initially have larger semi-major axes $a_i(S)$ than the most-bound element ($S = 0$) as indicated by Eq.~(\ref{E:ai}).  Although the initial self-intersection points are at comparable dimensionless distances from the SMBH for the three values of the SMBH mass $M_\bullet$ depicted in Fig.~\ref{fig:r_colVtime} (these three values straddle the maximum of the blue $\beta = 1$ curve in the left panel of Fig.~\ref{fig:r_col}), the tidal stream circularizes more rapidly for more massive SMBHs because the specific energy dissipated at the collision points scale as $v_t^2 \propto q^{-2/3}$ according to Eq.~(\ref{E:vt}).  The tidal stream eventually circularizes for all three values of $M_\bullet$ when the distance $r_{\rm col}(t)$ reaches the circularization radius $r_c$ given by Eq.~(\ref{E:rc}).

\begin{figure*}[t!]
\begin{subfigure}[t]{0.49\textwidth}
\includegraphics[width=\linewidth]{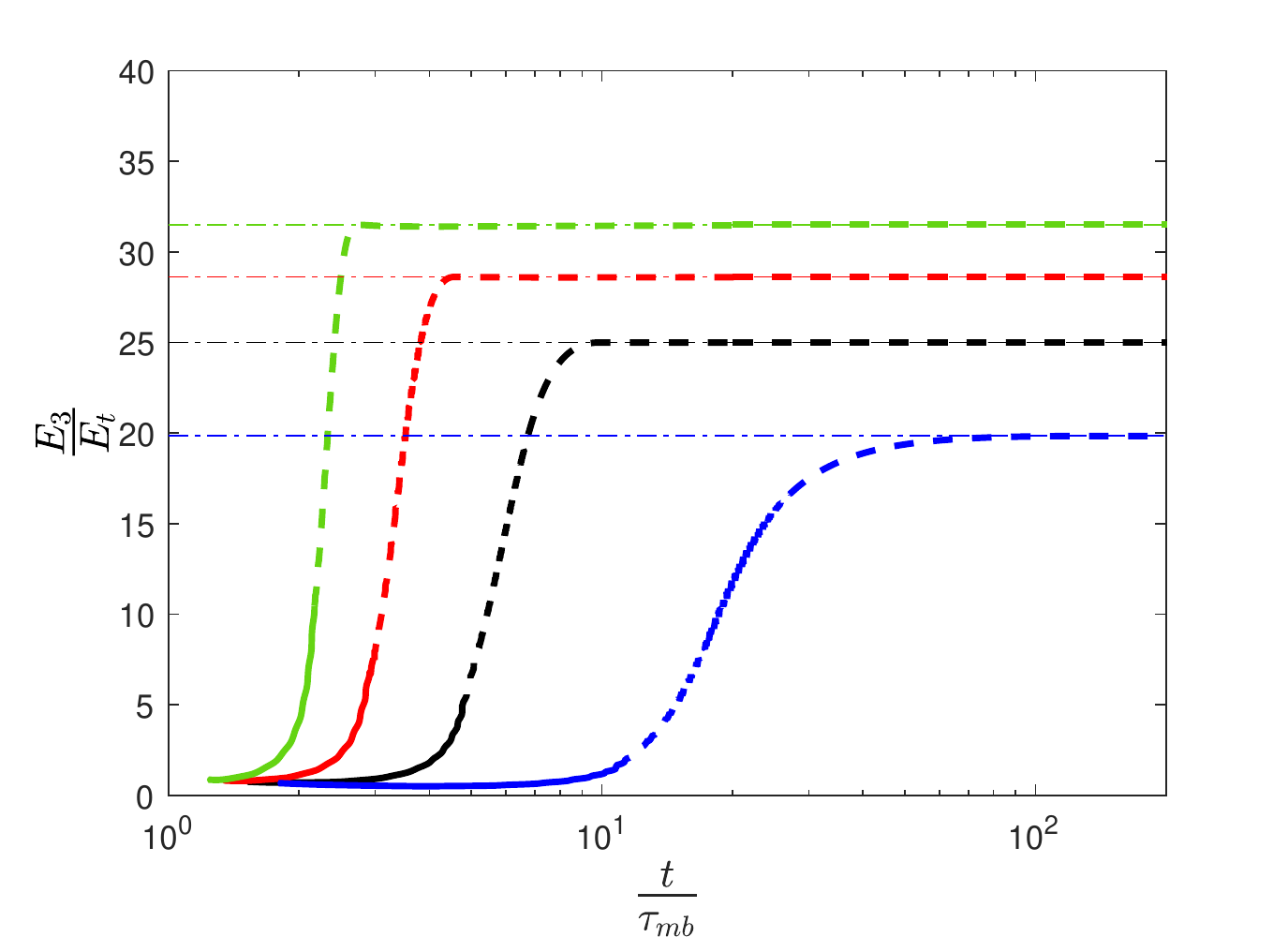}
\end{subfigure}
\begin{subfigure}[t]{0.49\textwidth}
\includegraphics[width=\linewidth]{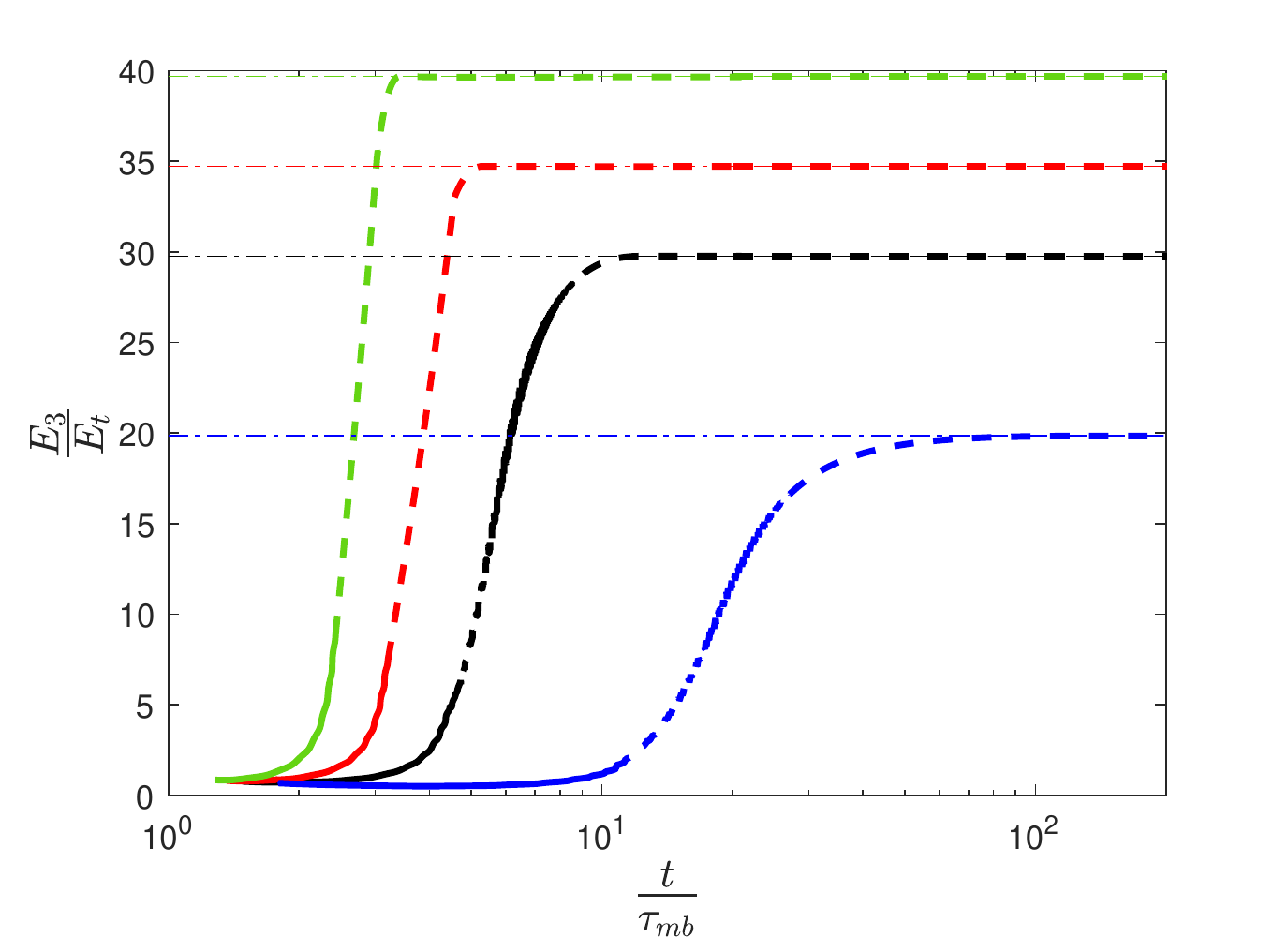}
\end{subfigure}
\caption{(Left panel) The specific binding energy $E_3$ of stream III as a function of time $t$ for TDEs with penetration factor $\beta = 1$ and SMBH masses $M_\bullet/M_\odot = 5 \times 10^5$ (blue), $10^6$ (black) $1.5 \times 10^6$ (red), and $2 \times 10^6$ (green).  (Right panel) The specific binding energy $E_3$ of stream III as a function of time $t$ for TDEs with SMBH mass $M_\bullet = 5 \times 10^5 M_\odot$ and penetration factors $\beta = 1$ (blue), $1.5$ (black) $1.75$ (red), and $2$ (green).  In both panels, the solid portions of the curves show kinematic simulations, the dashed portions show late-time extrapolations of these simulations, and the horizontal dotted lines show the circularization energy $E_c/E_t = \beta q^{-1/3}/4$.}
\label{F:SBE1}
\end{figure*}

Energy dissipation at the collision point causes stream III to become increasingly bound with time as shown in Fig.~\ref{F:SBE1}.  As the inelastic stream collisions conserve specific angular momentum, stream III becomes fully circularized when it reaches a specific binding energy
\begin{equation} \label{E:Ec}
E_c = \frac{GM_\bullet}{2r_c} = \frac{\beta}{4} q^{-1/3} E_t~.
\end{equation}
Although TDEs with higher SMBH masses $M_\bullet$ and penetration factors $\beta$ have higher circularization energies $E_c$, the greater apsidal precession given by Eq.~(\ref{E:PrecessionAngle}) implies that they circularize faster.  Our extrapolation (\ref{E:TA2fit}) of the true anomaly of stream II implies that the collision point 
\begin{equation} \label{E:rcolex}
r_{\rm col}(S_2) = \frac{a_i(1 - e_i)^2}{1 + e_i\cos f_2} 
\approx \frac{2r_c}{\tanh[k(S_2 - C)] + 1} 
\end{equation}
approaches the circularization radius $r_c$ by construction at late times ($S_2 \to \infty$), but it can take a very long time ($t_c \gg \tau_{\rm mb}$) to do so as indicated in Figs.~\ref{fig:r_colVtime} and \ref{F:SBE1} for $M_\bullet = 5 \times 10^5 M_\odot$ and $\beta = 1$.  If the breakdown of the "loop + tail" topology described in the Appendix had not necessitated such an extrapolation, our kinematic simulations might have demonstrated that energy dissipation at the collision point was too inefficient to circularize the tidal stream for such mild apsidal precession.

\begin{figure}[t!]
\includegraphics[width=\linewidth]{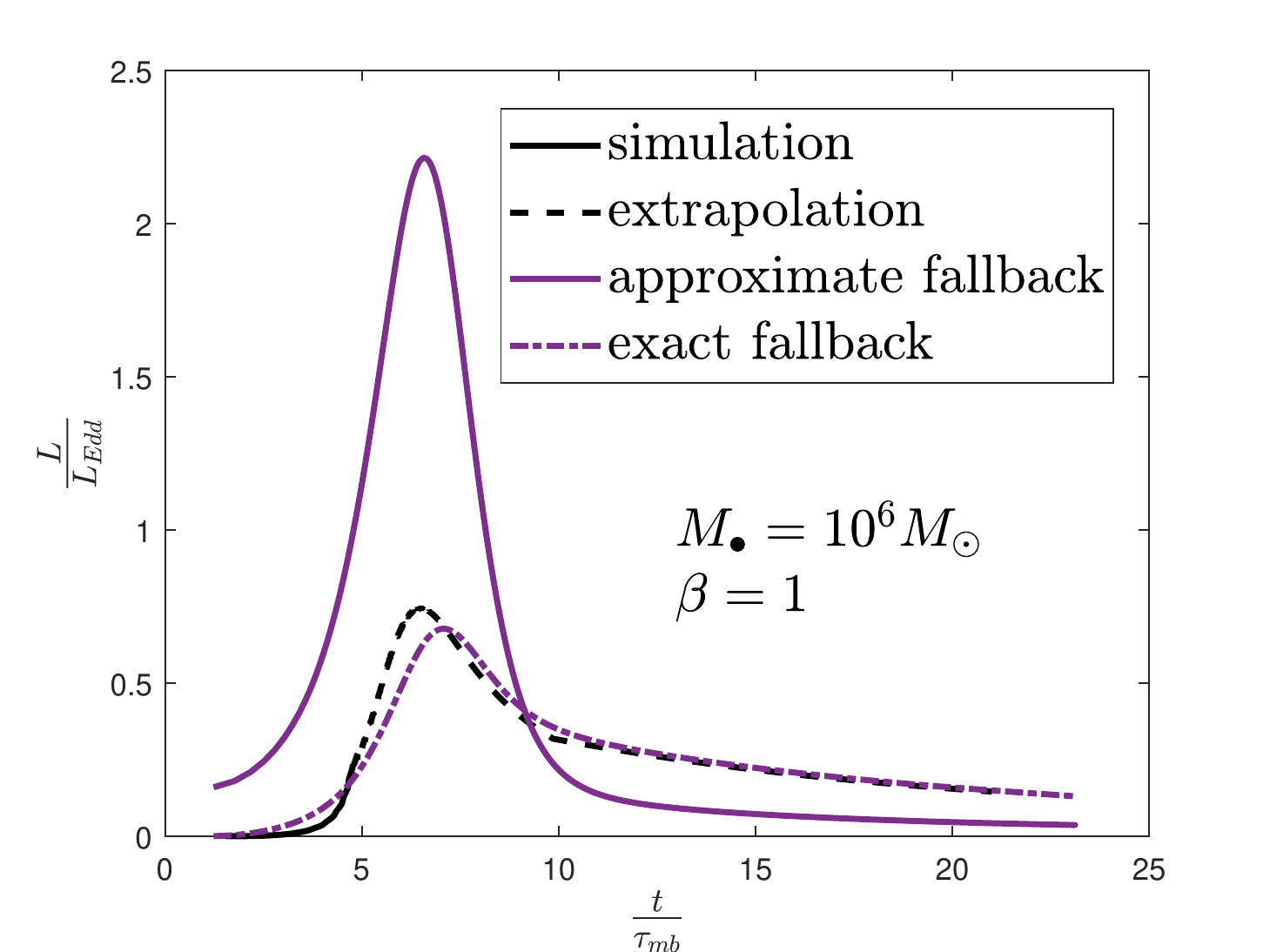}
\caption{Bolometric light curves associated with the energy dissipated in stream collisions for a TDE with SMBH mass $M_\bullet = 10^6M_\odot$ and penetration factor $\beta = 1$. The solid and dashed black curves are the predictions of our kinematic simulation and extrapolation, while the solid and dot-dashed purple curves are the predictions of \citeauthor{bonnerot17}~\cite{bonnerot17} using a flat energy distribution and the freeze-in model of \citeauthor{lodato09}~\cite{lodato09}.}
\label{fig:LCBonnerot}
\end{figure}

The bolometric luminosity associated with this energy dissipation at the collision point is shown in Fig.~\ref{fig:LCBonnerot} for our default choice of parameters $M_\bullet = 10^6 M_\odot$, $\beta = 1$.  Its rapid rise from zero at $t = t_{\rm col,1st}$ can be understood from Eq.~(\ref{E:lum}): the stream mass ratio $Q$, radial velocities $v_r(S_i)$, and mass distribution $dM/dS(S_2)$ all grow rapidly as the collision point approaches the circularization radius $r_c$.  It reaches a maximum of $L_{\rm peak} = 0.75L_{\rm Edd}$ at $t_{\rm peak} = 6.48\tau_{\rm mb}$ as the factor $Q/(1 + Q)$ asymptotes to unity, the radial velocity $v_r(S_1) \to 0$ as stream I circularizes, and $dM/dS(S_2)$ passes through its own maximum at $S_2 = S_{\rm max} = 4.78$ as shown in Fig.~\ref{fig:dm_dS}.  After stream I has circularized at $t_c = 9.85\tau_{\rm mb}$, the light curve traces the mass fallback rate as indicated by Eq.~(\ref{E:lumlatenorm}).  The relative proximity $t(S_{\rm max}) \simeq t_c$ of the two times leads to a significant enhancement of the peak bolometric luminosity $L_{\rm peak}$ above the circularized prediction of Eq.~(\ref{E:lumlatenorm}).

For comparison, we also show the predictions of the models of \citeauthor{bonnerot17}~\cite{bonnerot17} for the same choice of parameters, conserved angular momentum (Alfv\'{e}n velocity $v_A = 0$), and perfect radiative efficiency ($\eta = 1$).  The solid purple curve in Fig.~\ref{fig:LCBonnerot} shows the prediction of \citeauthor{bonnerot17}~\cite{bonnerot17} assuming an initially flat energy distribution of the tidal debris consistent with a mass distribution
\begin{equation} \label{E:flatdMdS}
\frac{dM}{dS} = \frac{dM}{dE_i} \left| \frac{dE_i}{dS} \right| = \frac{M_\star}{3} (1 + S)^{-5/3}. 
\end{equation}
This monotonically decreasing function vastly overestimates the mass distribution at early times compared to the more accurate freeze-in model of \citeauthor{lodato09}~\cite{lodato09} as can be seen by comparing it to $dM/dS$ shown in Fig.~\ref{fig:dm_dS}.  It also underestimates the mass distribution at late times ($S \to \infty$)  by a factor $\approx 0.226$ as can be seen in comparison with the limit given by Eq.~(\ref{E:dMdSLate}).  These discrepancies are largely responsible for the disagreements between the black and solid purple curves in Fig.~\ref{fig:LCBonnerot}.

The dot-dashed purple curve in Fig.~\ref{fig:LCBonnerot} shows the prediction of 
\citeauthor{bonnerot17}~\cite{bonnerot17} using the same initial mass distribution of \citeauthor{lodato09}~\cite{lodato09} adopted by our model. There is much greater agreement, although the \citeauthor{bonnerot17}~\cite{bonnerot17} model still overestimates (underestimates) the bolometric light curve at early (late) times $t < t_c$ prior to circularization.  This residual disagreement is primarily due to their assumption that the stream mass ratio $Q$ is always equal to unity in the "stream self-crossing shock luminosity" contribution $L^{\rm s}_{\rm sh}$ to the bolometric light curve.\footnote{This disagreement shown in Fig.~\ref{fig:LCBonnerot} is smaller than would be inferred from Fig.~8 of \citeauthor{bonnerot17}~\cite{bonnerot17}; we were unable to precisely reproduce their light curve despite our best efforts.}  Once stream I circularizes, $L^{\rm s}_{\rm sh} \to 0$ and the light curve is dominated by the "tail shock luminosity" $L^{\rm t}_{\rm sh}$ equal to the product of the mass fallback rate and the assumed homogeneous specific energy of stream I.  After circularization, $L^{\rm t}_{\rm sh}$ agrees with our late-time prediction $L_\infty$ of Eq.~(\ref{E:lumlate}).  One might expect greater disagreement between our model and that of \citeauthor{bonnerot17}~\cite{bonnerot17} for TDEs by smaller SMBH masses $M_\bullet$ whose streams circularize more slowly and have light curves that remain dominated by the "stream self-crossing shock luminosity" until later times.

\begin{figure*}[t!]
\begin{subfigure}[t]{0.49\textwidth}
\includegraphics[width=\linewidth]{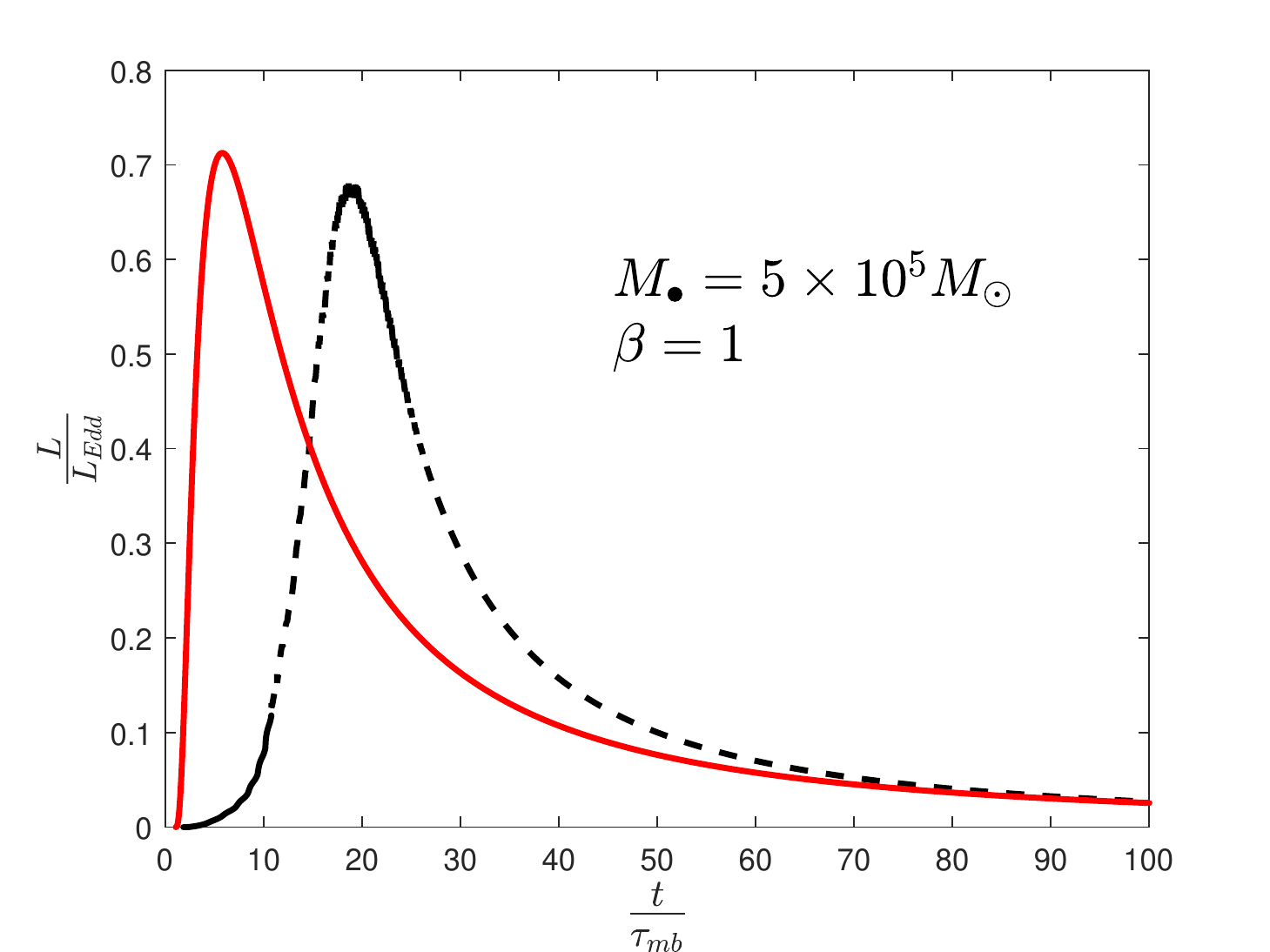}
\end{subfigure}
\begin{subfigure}[t]{0.49\textwidth}
\includegraphics[width=\linewidth]{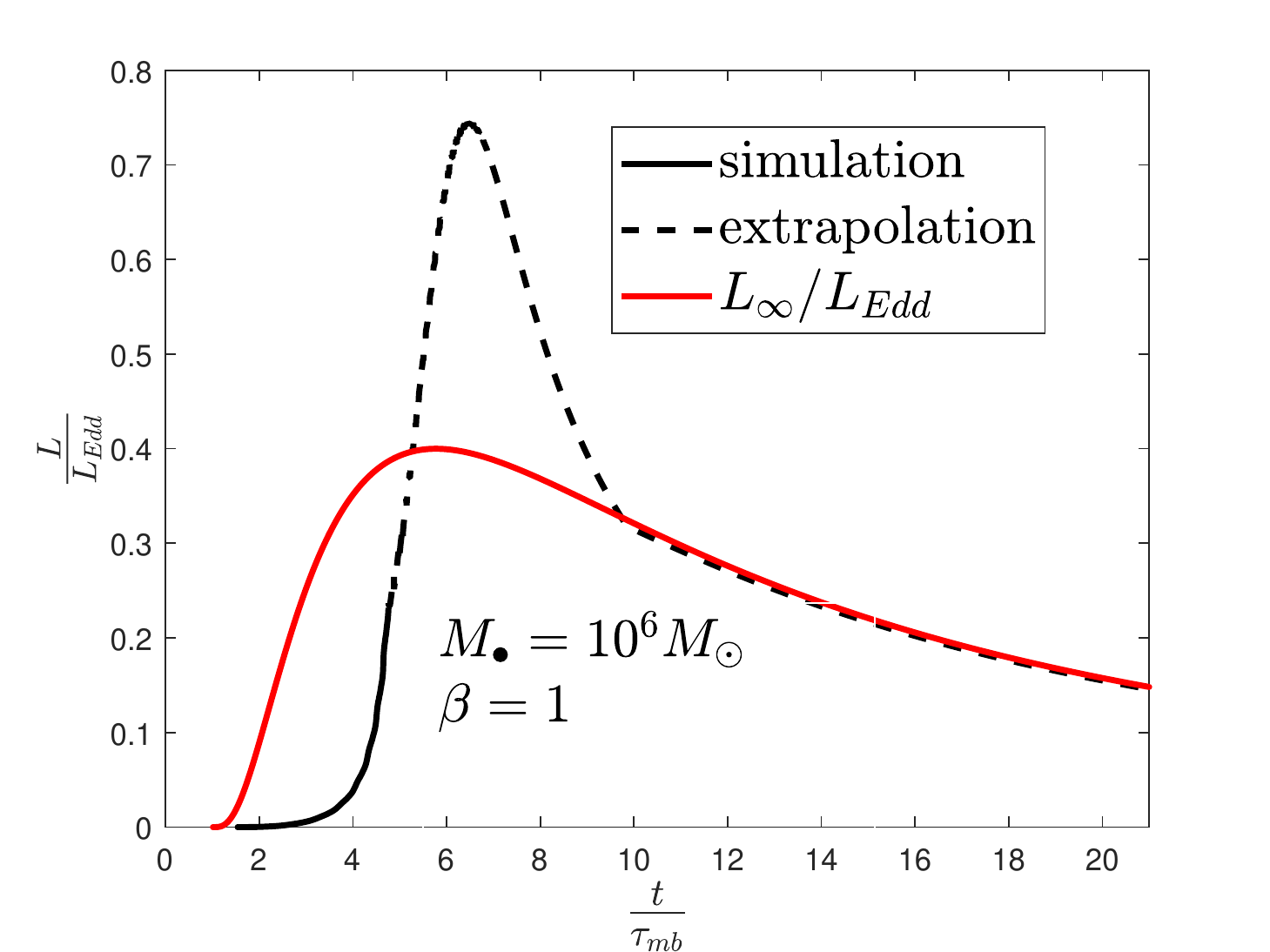}
\end{subfigure}
\begin{subfigure}[t]{0.49\textwidth}
\includegraphics[width=\linewidth]{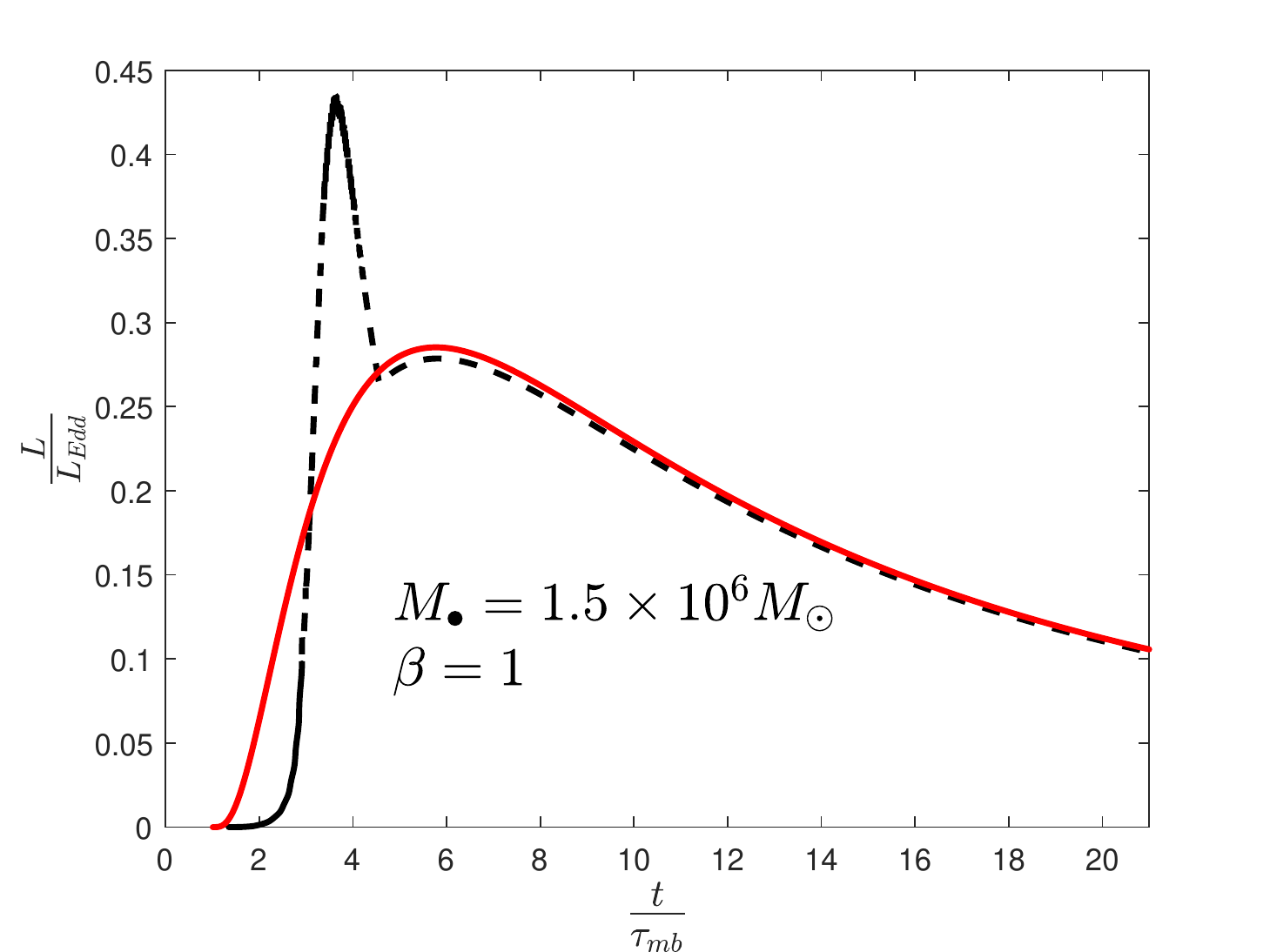}
\end{subfigure}
\begin{subfigure}[t]{0.49\textwidth}
\includegraphics[width=\linewidth]{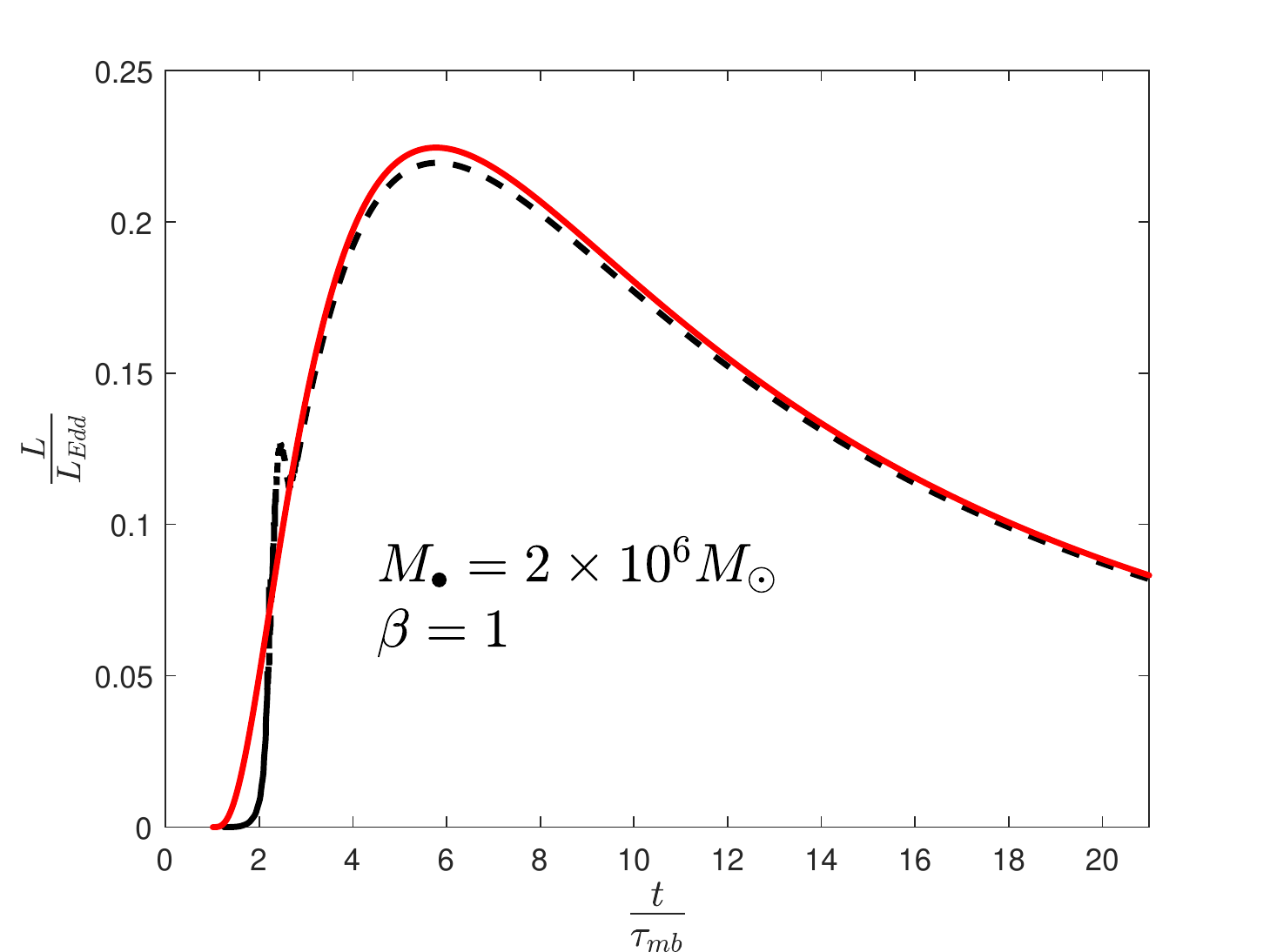}
\end{subfigure}
\caption{Bolometric light curves for TDEs by SMBHs with masses $M_\bullet/M_\odot = 5 \times 10^5$, $10^6$, $1.5 \times 10^6$, and $2 \times 10^6$.  The penetration factor is $\beta = 1$.  The solid (dashed) black curves depict our kinematic simulations (extrapolations) and the red curves show the late-time luminosity limit $L_\infty/L_{\rm Edd}$ of Eq.~(\ref{E:lumlatenorm}).}
\label{F:LCvsMass}
\end{figure*}

In Fig.~\ref{F:LCvsMass}, we explore the dependence of our predicted TDE bolometric light curves on SMBH mass $M_\bullet$ for fixed penetration factor $\beta = 1$.  The solid (dashed) black curves show the light curves given by our kinematic simulations (extrapolations) described in Secs.~\ref{SS:KS} and \ref{SS:extrap} respectively, while the solid red curves show the circularized bolometric luminosity $L_\infty/L_{\rm Edd}$ given by Eq.~(\ref{E:lumlatenorm}).  These latter curves are all proportional to the mass fallback rate $dM/dS$ shown in Fig.~\ref{fig:dm_dS} and reach a peak at $t_{\rm fb} = (1 + S_{\rm max})\tau_{\rm mb} = 5.78\tau_{\rm mb}$.  They are proportional to $M_\bullet^{-5/6}$, implying that they only predict super-Eddington peak luminosities for $M_\bullet \lesssim 3.3 \times 10^5 M_\odot$.  

Although our model agrees with the late-time limit $L_\infty$ after stream I circularizes ($t > t_c$), there are significant deviations at earlier times.  As the SMBH mass increases, the apsidal precession angle $\Delta\omega$ increases proportional to $M_\bullet^{2/3}$ as given by Eq.~(\ref{E:PrecessionAngle}).  This implies that the circularization time $t_c$ decreases as can be seen in Fig.~\ref{fig:r_colVtime} and the left panel of Fig.~\ref{F:SBE1}.  These circularization times are also listed in Table~\ref{T:circ} below.  For small SMBH masses, stream I circularizes after the peak in the mass fallback rate ($t_c > t_{\rm fb}$).  In this regime of "slow" circularization, considerable mass builds up in stream I, the "loop" of our "loop + tail" topology, before energy dissipation at the collision point is efficient enough to achieve circularization.  This implies a single peak in the bolometric light curve near the time at which the factor $f_v \equiv [v_r(S_1) - v_r(S_2)]^2$ in Eq.~(\ref{E:lum}) is maximized.  For the smallest SMBH masses, such as $M_\bullet = 5 \times 10^5 M_\odot$ shown in the top left panel of Fig.~\ref{F:LCvsMass}, the time $t_{\rm peak}$ at which this peak occurs can be much greater than the time $t_{\rm fb}$ at which the mass fallback peaks.  For larger SMBH masses for which $t_c$ is not much greater than $t_{\rm fb}$, such as $M_\bullet = 10^6 M_\odot$ shown in the top right panel of Fig.~\ref{F:LCvsMass}, the peaks in the factors $f_v$ and $dM/dS$ in Eq.~(\ref{E:lum}) can overlap.  This leads to a significant enhancement of the light curve ($\approx 90\%$ for $M_\bullet = 10^6 M_\odot$) above the circularized prediction $L_\infty$ given by Eq.~(\ref{E:lumlatenorm}).

In the opposite regime of "fast" circularization ($t_c < t_{\rm fb}$), larger apsidal precession $\Delta\omega$ leads to more efficient energy dissipation which circularizes stream I prior to the peak in the mass fallback rate at $t_{\rm fb}$.  This leads to two distinct peaks in the bolometric light curve: a first peak at which the velocity factor $f_v$ is maximized, and a second peak at $t_{\rm fb}$ when the mass fallback rate is maximized.  When $t_c$ is not much less than $t_{\rm fb}$, such as for $M_\bullet = 1.5 \times 10^6 M_\odot$ shown in the bottom left panel of Fig.~\ref{F:LCvsMass}, the maxima of the factors $f_v$ and $dM/dS$ in Eq.~(\ref{E:lum}) overlap and the first peak dominates over the second.  At the largest SMBH masses, such as for $M_\bullet = 2 \times 10^6 M_\odot$ shown in the bottom right panel of Fig.~\ref{F:LCvsMass}, stream I circularizes extremely rapidly ($t_c \ll t_{\rm fb}$), the first peak is merely a small fluctuation on the rising light curve, and the bolometric luminosity dissipated at the collision point traces the mass fallback rate consistent with Eq.~(\ref{E:lum}) for most of the duration of the TDE.  Note that this regime is still distinct from the canonical scenario of \citeauthor{rees88}~\cite{rees88}, in which the the bolometric luminosity is assumed to trace the mass fallback rate because the circularization time $t_c$ is much shorter than the viscous time at the circularization radius $r_c$.  Our model implies a radiative efficiency
\begin{align}
\eta_\infty &= \frac{E_c}{c^2} = \frac{\beta E_\star}{4q^{2/3}c^2} \notag \\
&= 5.3 \times 10^{-3} \beta \left(\frac{M_\bullet}{10^6M_\odot}\right)^{2/3} \left(\frac{M_\star}{M_\odot}\right)^{1/3} \left(\frac{R_\star}{R_\odot}\right)^{-1}
\end{align}
at late times, unlike the canonical scenario which predicts a radiative efficiency $\eta_{\rm ISCO} \approx 0.1$ consistent with the innermost stable circular orbit of the SMBH.

\begin{figure*}
\begin{subfigure}[t]{0.49\textwidth}
\includegraphics[width=\linewidth]{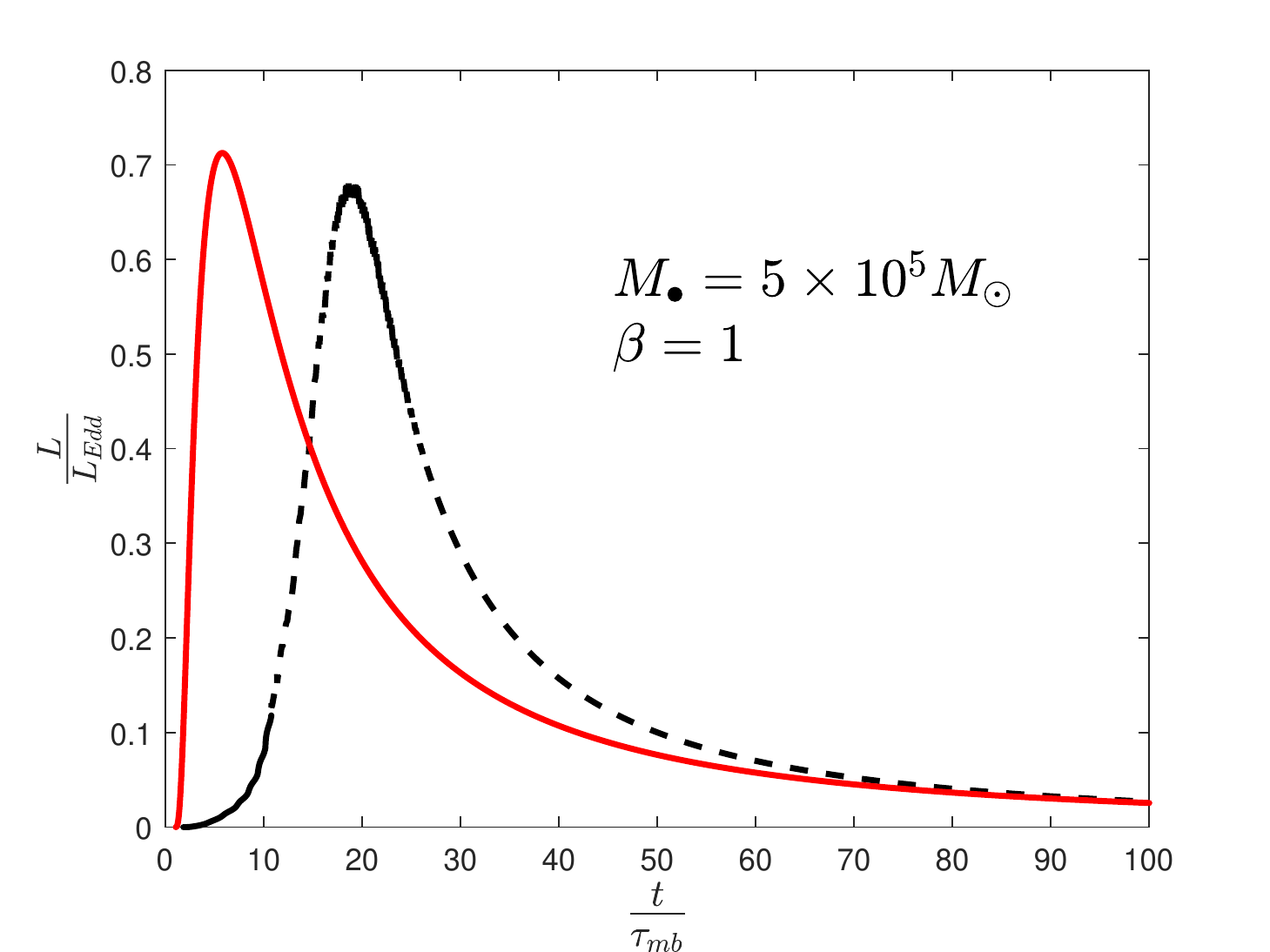}
\end{subfigure}
\begin{subfigure}[t]{0.49\textwidth}
\includegraphics[width=\linewidth]{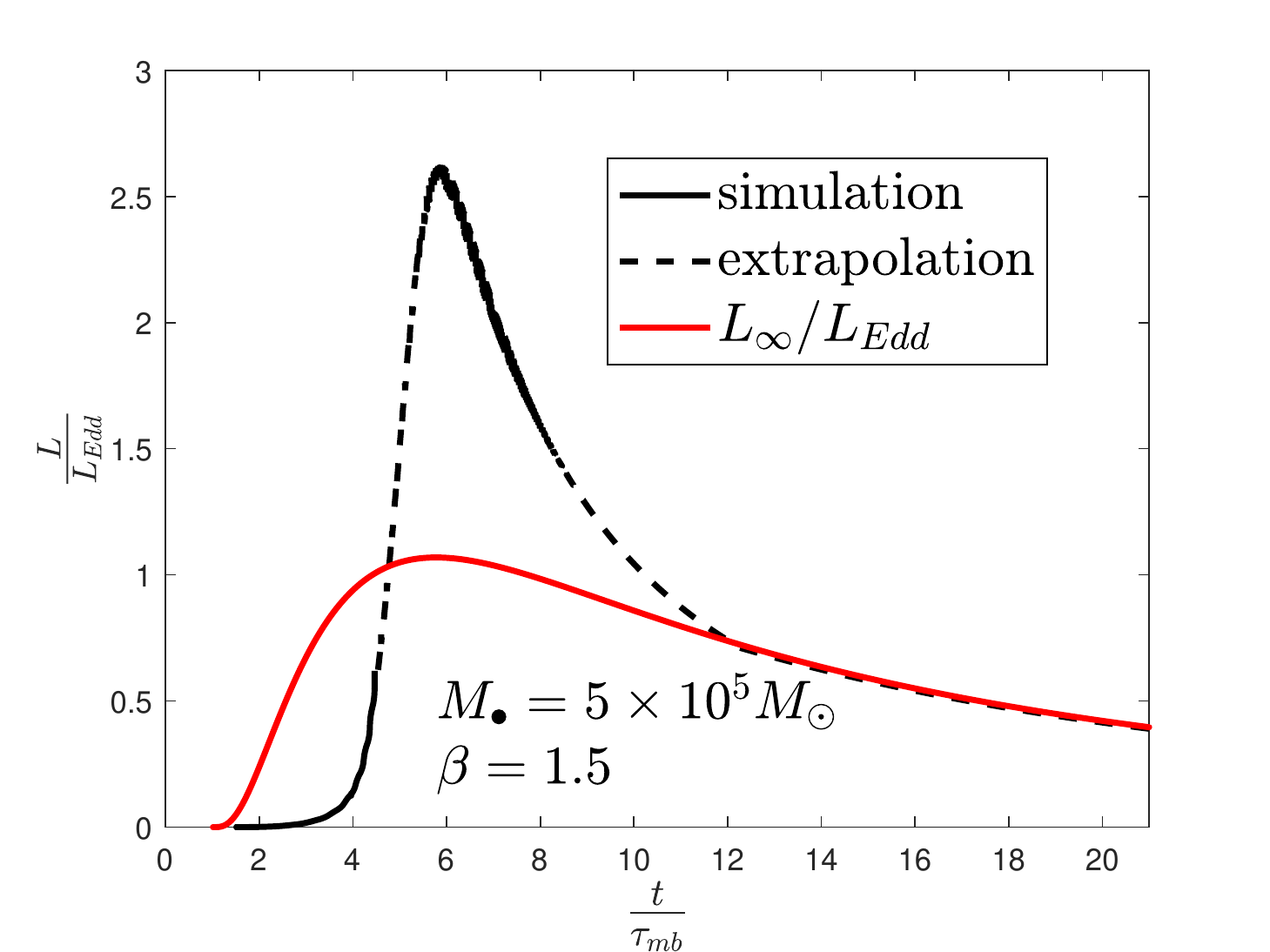}
\end{subfigure}
\begin{subfigure}[t]{0.49\textwidth}
\includegraphics[width=\linewidth]{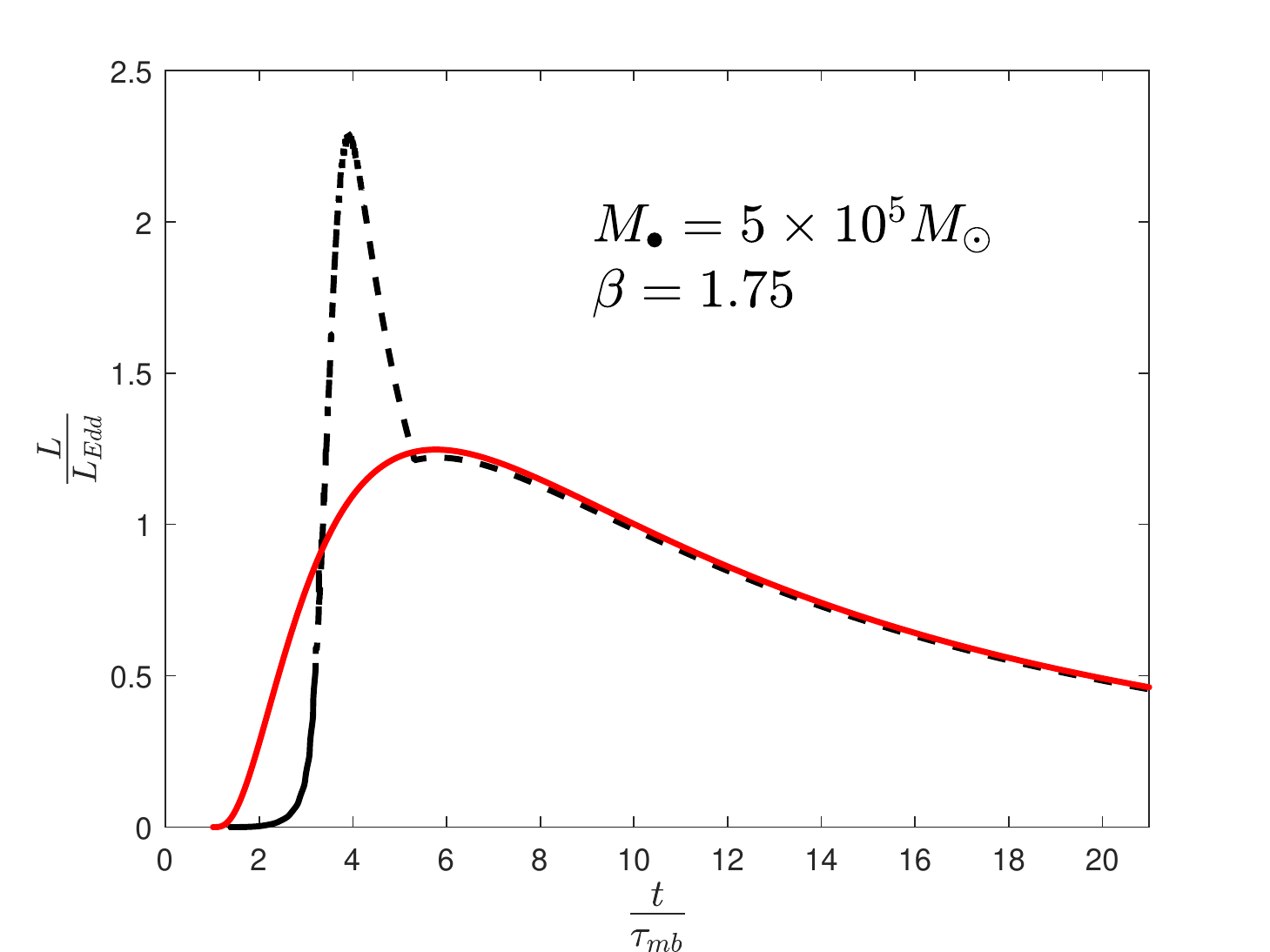}
\end{subfigure}
\begin{subfigure}[t]{0.49\textwidth}
\includegraphics[width=\linewidth]{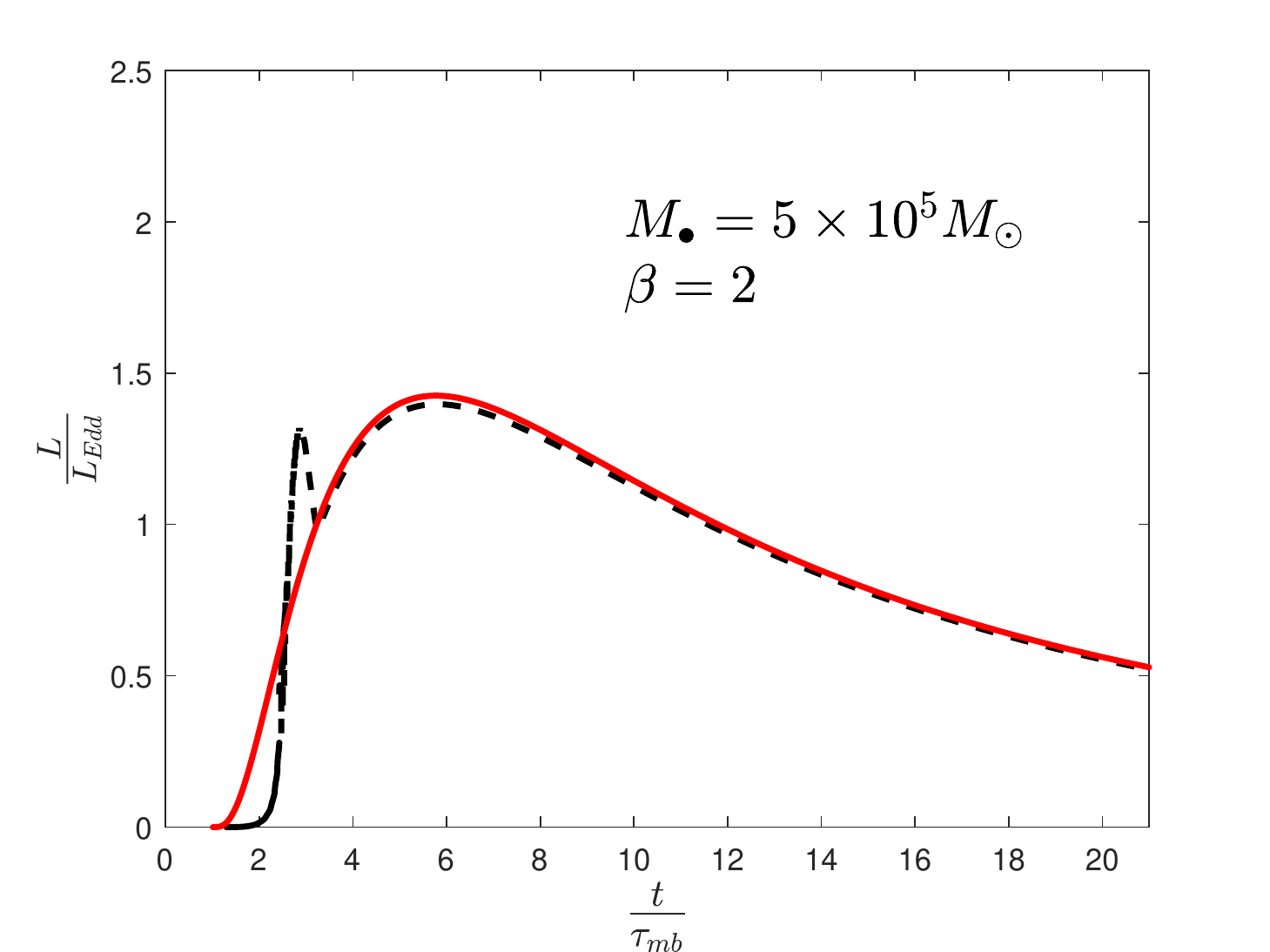}
\end{subfigure}
\caption{Bolometric light curves for TDEs with penetration factors $\beta = 1$, $1.5$, $1.75$, and $2$.  The SMBH mass is $M_\bullet = 5 \times 10^5 M_\odot$.  The solid (dashed) black curves depict our kinematic simulation (extrapolations), while the red curves show the late-time luminosity limit $L_\infty/L_{\rm Edd}$ of Eq.~(\ref{E:lumlatenorm}).}
\label{F:LCvsBeta}
\end{figure*}

In Fig.~\ref{F:LCvsBeta}, we examine how our bolometric light curves depend on the penetration factor $\beta$ for a fixed SMBH mass $M_\bullet = 5 \times 10^5 M_\odot$.  The top left panel of this figure is identical to that of Fig.~\ref{F:LCvsMass}, but the other panels appear qualitatively similar as well.  This is primarily because the apsidal precession angle of Eq.~(\ref{E:PrecessionAngle}) that determines the efficiency of circularization is also a monotonically increasing function of the penetration factor ($\Delta\omega \propto \beta$).  The overall normalization of these light curves is different however, as according to Eq.~(\ref{E:lumlatenorm}) the circularized bolometric lumnosity increases with penetration factor ($L_\infty/L_{\rm Edd} \propto \beta$) while it decreases with SMBH mass ($L_\infty/L_{\rm Edd} \propto M_\bullet^{-5/6}$).  As the penetration factor $\beta$ increases, the circularization time $t_c$ rapidly decreases.  The circularization regime transitions from "slow" to "fast" for some penetration factor $1.5 \leq \beta_{\rm trans}(M_\bullet) \leq 1.75$, and by $\beta = 2$ the light curve traces the mass fallback rate even for this comparatively small SMBH mass.

\begin{table*}[t!]
    \centering
    \begin{tabular}{|c|c|c|c|c|c|c|}
    \hline
    $M_\bullet/M_\odot$ & $\beta$ & $\Delta \omega$ & $t_{\rm peak} / \tau_{\rm mb}$ & $t_c / \tau_{\rm mb}$ & $L_{\rm peak}/L_{\rm Edd}$ & Circularization \\
    \hline
    $5 \times 10^5$ & 1 & 0.126 & 17.91 & 141.14 & 0.68 & Slow \\
    $5 \times 10^5$ & 1.5 & 0.189 & 5.80 & 8.58 & 2.21 & Slow \\
    $7 \times 10^5$ & 1.2 & 0.189 & 6.07 & 9.29 & 1.38 & Slow \\
    $10^6$ & $1$ & 0.2 & 6.48 & 9.85 & 0.75 & Slow \\
    \hline
    $5 \times 10^5$ & 1.75 & 0.220 & 3.91 & 5.31 & 2.29 & Fast \\
    $5 \times 10^5$ & 2 & 0.252 & 2.87 & 3.36 & 1.32 & Fast \\
    $1.5 \times 10^6$ & 1 & 0.262 & 3.65 & 4.56 & 0.43 & Fast \\
    $10^6$ & $1.5$ & 0.3 & 2.19 & 2.33 & 0.30 & Fast \\
    $2 \times 10^6$ & 1 & 0.318 & 2.43 & 2.81 & 0.13 & Fast \\
    \hline
    \end{tabular}
    \caption{SMBH mass $M_\bullet$, penetration factor $\beta$, apsidal precession angle $\Delta\omega$, time $t_{\rm peak}$ of the first luminosity peak, time $t_c$ at which stream I circularizes, luminosity $L_{\rm peak}$ of the first luminosity peak, and circularization regime.  Circularization is slow (fast) if $t_c$ is greater (less) than the time $t_{\rm fb} = 5.78\tau_{\rm mb}$ at which the mass fallback rate peaks.
    } \label{T:circ}
\end{table*}

In Table~\ref{T:circ}, we summarize the results of the simulations depicted in Figs.~\ref{F:LCvsMass} and \ref{F:LCvsBeta}, along with two additional simulations.  Although the circularization time $t_c$ is not a monotonically decreasing function of the apsidal precession angle $\Delta\omega$ for all choices of the parameters, all of the TDEs in the "fast" regime have $\Delta\omega \gtrsim 0.2$.  If we use this criterion as a crude estimate of the boundary between the "slow" and "fast" circularization regimes, then according to Eq.~(\ref{E:PrecessionAngle}), TDEs with penetration factors
\begin{equation} \label{E:betatrans}
\beta \gtrsim \beta_{\rm trans}(M_\bullet) = \left(\frac{M_\bullet}{10^6M_\odot}\right)^{-2/3}
\end{equation}
circularize promptly.  TDEs deep within the fast regime ($\beta \gg \beta_{\rm trans}(M_\bullet)$) will have bolometric light curves $L_\infty(t)$ given by Eq.~(\ref{E:lumlate}) that trace the mass fallback rate as in the bottom right panels of Figs.~\ref{F:LCvsMass} and \ref{F:LCvsBeta}.

\section{Discussion} \label{S:disc}

A star that passes sufficiently within the tidal radius $r_t$ of a SMBH will inevitably be disrupted, but the luminsoity and timescale of the electromagnetic emission by the resulting tidal debris is highly uncertain.  In the absence of relativistic apsidal precession, elements of the tidal stream will have different semi-major axes and eccentricities but will share a common argument of pericenter.  Whether such a highly eccentric tidal stream could successfully develop a magneto-rotational instability \cite{1991ApJ...376..214B} that would transform it into an accretion disk about the SMBH is unclear.  Fortunately for astronomers, relativistic apsidal precession causes the tidal stream to intersect with itself, leading to continuous inelastic collisions between a loop of circularizing debris (stream I) and a tail (stream II) falling back to pericenter for the first time.

Global hydrodnamic simulations of tidal disruption and stream circularization are extraordinarily computationally expensive given the huge dynamical range of time and length scales (see Sec.~3.1 of \citeauthor{2019GReGr..51...30S}~\cite{2019GReGr..51...30S} for a brief review of these numerical challenges).  This computational expense suggests that the one-dimensional kinematic simulations of the model presented in this paper can be useful to gain qualitative insight into tidal-stream circulation and more rapidly explore the parameter space.  Our model would need to be supplemented with three-dimensional radiation hydrodynamic simulations such as those presented in \citeauthor{2016ApJ...830..125J}~\cite{2016ApJ...830..125J} to quantitatively predict the radiative efficiency and effective temperature of these stream collisions.  These simulations indicate that stream collisions can indeed produce much of the optical/UV emission associated with TDEs, which might help to explain observed delays between optical/UV and X-ray emission \cite{pasham17,gezari17}.

The primary finding of our study, shown in Figs.~\ref{F:LCvsMass} and \ref{F:LCvsBeta} and summarized in Table~\ref{T:circ}, is that tidal-stream circularization can be divided into slow and fast regimes depending on whether the time $t_c$ on which the debris within the loop circularizes is longer or short than the time $t_{\rm fb}$ at which the mass fallback rate peaks.  In the slow regime, the mass contained in the fallback peak assembles into an eccentric loop that slowly radiates in a single broad peak as it circularizes.  In the fast regime, the eccentric loop rapidly circularizes producing a first narrow peak in the bolometric light curve.  This is followed by a second broader peak as the mass fallback rate of the tidal tail (stream II) onto the already circularized loop passes through its own peak at $t_{\rm fb}$.  TDE circularization occurs in the fast regime for apsidal precession angles $\Delta\omega \gtrsim 0.2$ which occur for penetration factors $\beta \gtrsim \beta_{\rm trans}(M_\bullet) = (M_\bullet/10^6M_\odot)^{-2/3}$. 

This criterion most likely overestimates the efficiency of tidal-stream circularization (underestimates $\beta_{\rm trans}$) for several reasons: (1) stream collisions are unlikely to be fully inelastic as assumed in our model, (2) the extrapolation beyond the breakdown of our kinematic simulations assumed that the stream circularizes exponentially, and (3) we have neglected nodal precession that can delay the first stream self-intersection \cite{2015ApJ...809..166G}.  If circularization is so inefficient that it fails entirely, the TDE may fail to produce an observable flare in the UV/optical or X-ray.  The rate $\dot{N}$ of TDEs by intermediate-mass black holes (IMBHs) with masses $M_\bullet \lesssim 10^6 M_\odot$ is dominated by the full loss-cone (pinhole) regime in which the distribution of penetration factors is $d\dot{N}/d\beta \propto \beta^{-2}$ \cite{2016MNRAS.455..859S,2020SSRv..216...35S}.  This implies that only a small faction of TDEs by IMBHs may be observable, which would help to reconcile the observed distribution of TDE host galaxies (for which $10^6 M_\odot \lesssim M_\bullet \lesssim 10^7 M_\odot$ is inferred from the $M_\bullet-\sigma$ relation \cite{2000ApJ...539L...9F,2000ApJ...539L..13G}) with theoretical predictions that the TDE rate would be dominated by dwarf galaxies hosting IMBHs \cite{2004ApJ...600..149W}.  Although it is challenging to make more specific observational predictions without radiative simulations, the identification of TDEs with double-peaked bolometric light curves associated with the transition between the slow and fast regimes of stream circularization by future high-cadence optical/UV surveys, like the Legacy Survey of Space and Time (LSST) by the Vera Rubin Observatory, would be a powerful indication that our model describes a key feature of TDE dynamics.

\acknowledgements

The authors were supported by NASA award number 80NSSC18K0639.  M.K. would like to thank the organizers and participants of the international workshop "Tidal Disruption Events: General Relativistic Transients" held at the Yukawa Institute for Theoretical Physics (YITP), Kyoto University on January 14 - 15, 2020 for valuable conversations, and the YITP for its hospitality and support during this workshop.

\appendix*

\section{Model Breakdown} \label{sec:appendix}

\begin{figure}
\centering
\includegraphics[width = 0.49\textwidth]{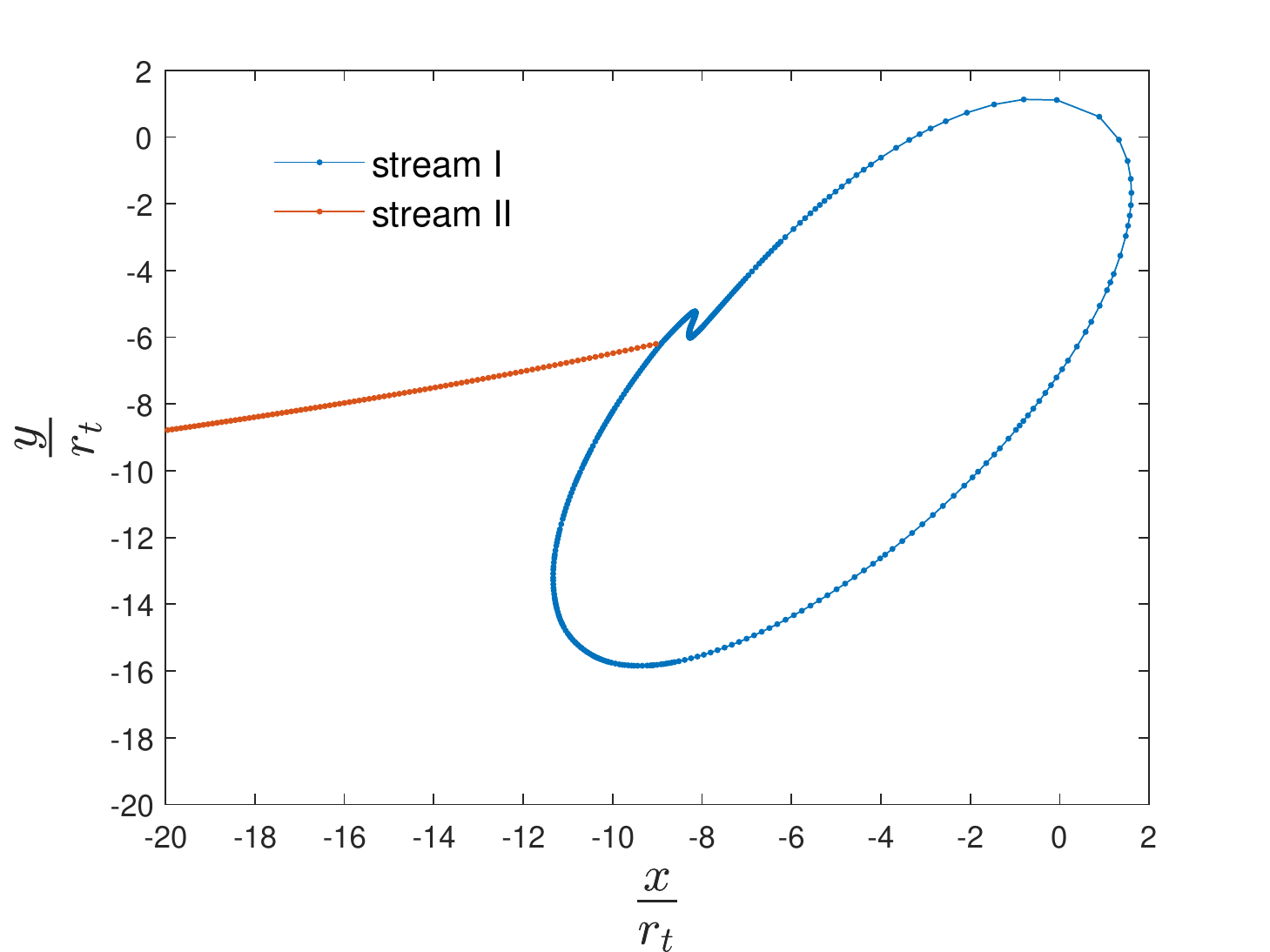}
\caption{Snapshot of the tidal stream at $t = 4.82 \tau_{\rm mb}$ for $M_\bullet = 10^6 M_\odot$ and $\beta = 1$.  Beyond this time, the kink in stream I destroys the simple "loop + tail" topology essential to our model of stream circularization.}
\label{F:kink}
\end{figure}

\begin{figure}
\centering
\includegraphics[width = 0.49\textwidth]{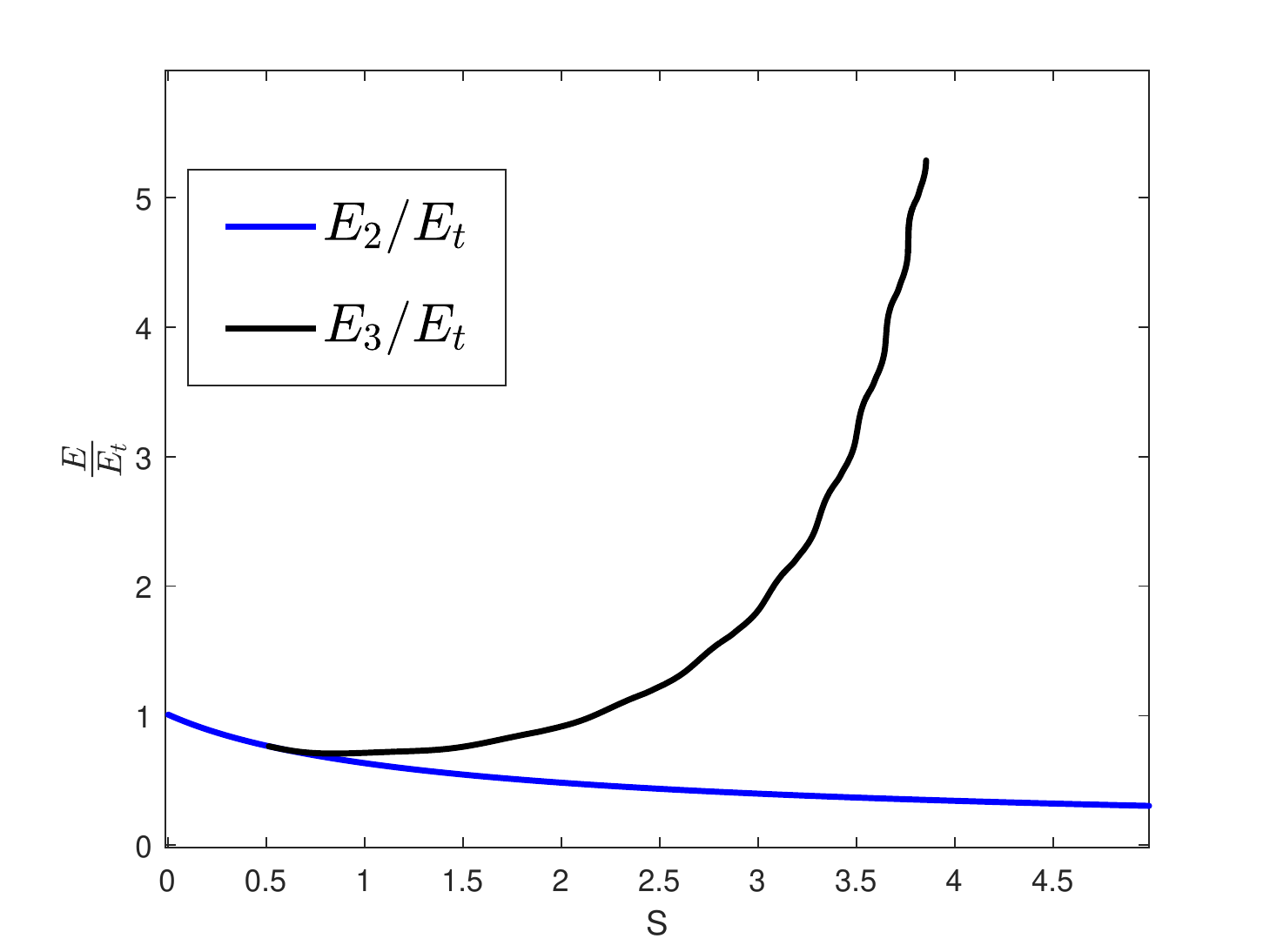}
\caption{Specific binding energy $E$ as a function of dimensionless fallback time $S$ for $M_\bullet = 10^6 M_\odot$ and $\beta = 1$.  The blue and black curves show stream II and III.  The curve for stream III ends at $t = 4.82 \tau_{\rm mb}$.
}
\label{F:EvsS}
\end{figure}

Our kinematic model of tidal stream circularization described in Sec.~\ref{SS:KS} is predicated on the stream maintaining the "loop + tail" topology pictured in the top left panel of Fig.~\ref{fig: stream reprocessing}.  This topology allows us to determine the element $S_1$ of stream I that collides with infalling element $S_2$ of stream II and thus use Eqs.~(\ref{E:MassCon}) and (\ref{E:MomCon}) to calculate the mass and velocity of the newly formed element $S_3 = S_2$ of stream III.  However, while performing these simulations, we discovered the emergence of a kink in stream I that eventually prevented the identification of a unique collision point.  Fig.~\ref{F:kink} shows a snapshot of the tidal streams for our default parameters $M_\bullet = 10^6 M_\odot$ and $\beta = 1$ at $t = 4.82 \tau_{\rm mb}$, shortly before the kink causes our model to break down.

Although we assumed at first that this kink was a numerical artifact, its existence and location were robust to changes in our numerical resolution.  After careful investigation, we concluded that it was a consequence of the inversion of specific binding energy in the loop which allows trailing elements to overtake leading elements before they return to the collision point.

The tidal stream is created at $t = 0$ when the tidally disrupted star first crosses the tidal radius $r_t$.  All the bound elements $S$ are within a stellar diameter $2R_\star$ at that time, but the distribution of initial specific binding energy $E_i(S) = (1 + S)^{-2/3}E_t$ given by Eq.~(\ref{E:Ei}) causes the less tightly bound elements (with longer orbital periods) to trail the more tightly bound (shorter period) leading elements.  This standard ordering of the specific binding energy for tidal streams is expressed in our notation by a monotonically decreasing function $E(S)$ such as that for stream II $E_2(S_2) = E_i(S_2)$ which has yet to experience its first inelastic collision.  This function is shown by the blue curve in Fig.~\ref{F:EvsS}.  This ordering is initially preserved because the stream mass ratio $Q = 0$ at the initial stream self-intersection implies that no energy is dissipated at $t = t_{\rm col,1st}$.  However, as $Q$ increases, increasing energy dissipation at the collision point causes the specific binding energy $E_3(S_3 = S_2)$ to reach a minimum at $S_{\min} = 0.84$ and then monotonically increase with time.  If this curve becomes sufficiently steep, trailing elements will have high enough binding energies (short enough periods) to overtake the leading elements before they make it around the loop and return to the collision point.  We can crudely estimate the criterion for this steepness as $d\tau/dt \lesssim -1$ which implies
\begin{equation} \label{E:breakdown}
\frac{d(E_3/E_t)}{dS} \gtrsim \frac{2}{3} \left( \frac{E}{E_t} \right)^{5/2}.
\end{equation}
This criterion is satisfied by the black curve $E_3(S)/E_t$ in Fig.~\ref{F:EvsS} before $t = 4.82 \tau_{\rm mb}$ when the leading edge of the kink shown in Fig.~\ref{F:kink} overtakes its trailing edge before reaching the collision point.

In a genuine hydrodynamics simulation, the tidal streams will have finite thickness and the formation of shocks will prevent kinks like those that cause our kinematic simulations to break down.  The delayed "Shock 2b" formed in the general relativistic hydrodynamics simulations of \citeauthor{shiokawa15}~\cite{shiokawa15} and depicted in the right panel of their Fig.~12 may result from an inversion in the ordering of the specific energy in the "loop" like described above.  If hydrodynamic forces are able to smooth out the kink without disrupting the "loop + tail" morphology of the tidal stream, it is possible that the kinematic approach of our model can still qualitatively describe the evolution of the tidal stream all the way to circularization.  This assumption motivates the extrapolation described in Sec.~\ref{SS:extrap} that replaces the conservation of linear momentum given by 
Eqs.~(\ref{E:MomCon}) with the fits of Eqs.~(\ref{E:TA2fit}) and (\ref{E:S1S2fit}) to model stream circularization beyond the breakdown in our kinematic simulations.

\bibliography{main}

\end{document}